\documentclass[useAMS,usenatbib]{mn2e}
\usepackage{rotating}
\usepackage{longtable}
\usepackage{graphics}
\usepackage{times}

\def\teff{$T_{\rm eff}$}
\def\lgg{$\log g$}

\def\vs{$v_{\rm e}\sin i$}

\def\pr{Pr\,{\sc iii}}
\def\nd{Nd\,{\sc iii}}
\def\dy{Dy\,{\sc iii}}
\def\ha{H$\alpha$}

\newcommand{\bs}{$\langle B \rangle$}
\def\kms{km\,s$^{-1}$}
\def\ms{m\,s$^{-1}$}

\newcommand{\figps}[1]{\resizebox{\hsize}{!}{\rotatebox{0}{\includegraphics{#1}}}}

\newcommand{\fifps}[2]{\centering\resizebox{#1}{!}{\includegraphics{#2}}}

\def\i{\,{\sc i}} \def\ii{\,{\sc ii}} \def\iii{\,{\sc iii}}

\begin{document}

\title[Pulsations in the roAp star 10 Aql]
{Pulsations in the atmosphere of the rapidly oscillating Ap star 10\,Aquilae\thanks{Based on observations collected at the European Southern Observatory, Paranal, Chile(program 077.D-0491 and a program 077.D-0150, retrieved through the ESO archive) and at the Telescopio Nazionale Galileo (TNG).}}
\author[M.~Sachkov et al.]
{M.~Sachkov$^1$\thanks{E-mail: msachkov@inasan.ru},
O.~Kochukhov$^2$, T.~Ryabchikova$^{1,3},$
D.~Huber$^3$, F.~Leone$^4$,
S.~Bagnulo$^5$ 
\newauthor 
and W.W.~Weiss$^3$ \\
$^1$ Institute of Astronomy, Russian Academy of Sciences, Pyatnitskaya 48, 119017 Moscow, Russia \\
$^2$ Department of Physics and Astronomy, Uppsala University Box 515, SE-751 20 Uppsala, Sweden\\
$^3$ Department of Astronomy, University of Vienna, T\"urkenschanzstrasse 17, A-1180 Wien, Austria\\
$^4$ Dipartimento di Fisica ed Astronomia, Universit\'a di Catania, via S. Sofia 78, 95123 Catania, Italy\\
$^5$ Armagh Observatory, College Hill, Armagh BT61 9DG, Northern Ireland, UK}

\date{Accepted 2008. Received 2008; in original form 2008}

\pagerange{\pageref{firstpage}--\pageref{lastpage}}
\pubyear{2008}

\maketitle

\label{firstpage}

\begin{abstract}
The rapidly oscillating Ap (roAp) star 10\,Aql shows one of the lowest photometric pulsation amplitudes
and is characterized by an unusual spectroscopic pulsational behavior compared to other roAp stars.
In summer 2006 this star became target of an intense observing campaign, that combined ground-based
spectroscopy with space photometry obtained with the MOST satellite.
More than 1000 spectra were taken during 7 nights over a time span of 21 days with 
high-resolution spectrographs at the 8-m ESO VLT and 3.6-m TNG telescopes giving access to
radial velocity variations of about 150 lines from different chemical species. A comparison of pulsation
signatures in lines formed at different atmospheric heights allowed us to resolve the vertical structure
of individual pulsation modes in 10\,Aql\ which is the first time for a multiperiodic roAp star. 
Taking advantage of the clear oscillation patterns seen in a number of rare-earth ions and using the
contemporaneous MOST photometry to resolve aliasing in the radial velocity measurements,
we improve also the determination of pulsation frequencies.
The inferred propagation of pulsation waves in 10\,Aql is qualitatively similar to other
roAp stars: pulsation amplitudes become measurable in the layers where Y and Eu are concentrated, increase in layers where the 
H$\alpha$ core is formed, reach a maximum of 200--300\,\ms\ in the layers probed by Ce, Sm, Dy lines and then
decrease to 20--50\,\ms\ in the layers where \nd\ and \pr\ lines are formed.
A unique pulsation feature of 10\,Aql is a second pulsation maximum indicated by Tb\iii\
lines which form in the uppermost atmospheric layers and oscillate with amplitudes of up to 350\,\ms.
The dramatic decline of pulsations in the atmospheric layers probed by the
strong \pr\ and \nd\ lines accounts for the apparent peculiarity of 10\,Aql when
compared to other roAp stars. The phase-amplitude diagrams and bisector measurements of the \nd\ 5102\,\AA\ line reveal
a rapid change of phase and amplitude with height for all three main pulsation modes, indicating the presence of a
pulsation node in the stellar atmosphere.
Finally, we report the discovery of a puzzling asymmetry of the strong \nd\ lines with their blue wing extending 
up to $-50$\,\kms, which is about 25 times the estimated value of \vs.
\end{abstract}


\section{Introduction}
\label{intro}

More than 30 members among the group of late A magnetic chemically peculiar stars exhibit high-overtone, low-degree,
non-radial $p$-mode pulsations with periods in the range of 6--21 minutes \citep{KM00}. These
\textit{rapidly oscillating Ap} (roAp) stars are characterized by strong global magnetic fields with a polar
strengths of the order of 1--10 kG.  The atmospheres of roAp stars are enriched with heavy elements, brought  up 
from the stellar interior by diffusion. Conspicuous lateral and vertical variations of chemical abundances
in Ap stars and the prominence of the magnetic field signatures make them primary targets for detailed
investigations of the magnetic topology and magnetically driven formation of structures in stellar atmospheres
(\citealt{K04a}; \citealt{R04}).

The observed pulsational amplitudes of roAp stars are modulated according to the visible magnetic field structure.
This observation led to the oblique pulsator model \citep{k1982}, where axisymmetric $\ell=1$ modes are
aligned with the magnetic field axis, which itself is oblique to the
axis of stellar rotation. Calculations by \citet{SG04} showed that dipolar modes
are significantly distorted by a magnetic field of kG-strength, but retain their axisymmetric
character. On the other hand, the structure of pulsational perturbations can include significant non-axisymmetric
components in weak-field stars \citep{BD02}. The indirect pulsation Doppler imaging of
oscillations in the prototype roAp stars HR\,3831 \citep{K04b} vindicated the oblique pulsator model and
for the first time characterized observationally the magnetic distortion of the global {\it p-}modes.

The roAp stars are key objects for asteroseismology, which presently is the most powerful tool for testing theories
of stellar structure and evolution. The classical asteroseismic analysis, utilizing precise frequency
measurements, helps to constrain the luminosity and internal chemical composition of pulsating magnetic
stars (e.g. Matthews, Kurtz \& Martinez \citeyear{MKM99}; Cunha, Fernandes \& Monteiro \citeyear{CFM03}).
Recent time-resolved spectroscopic observations of roAp stars uitilizing large telescopes  (see review by \citealt{K07a}) 
demonstrated the possibility of another type of asteroseismic investigation, which is focused on the upper
atmospheric layers. A significant chemical stratification and a short vertical wavelength of pulsation
modes lead to a remarkable diversity of the pulsational characteristics observed in individual spectral lines,
notably of absorption lines of the rare-earth elements (REE) (e.g., \citealt{KR01};
Mkrtichian, Hatzes \& Kanaan \citeyear{MHK03}; \citealt{ryab2007a}). The information extracted from the lines formed at different
optical depths opens access to different modes and can be combined to yield a vertical tomographic map of the
pulsating stellar atmosphere \citep{ryab2007b}. Thus, applying Doppler
imaging techniques and pulsation tomography to line profile variation in roAp stars promises an
unprecedented 3-D picture of  magnetoacoustic pulsations \citep{K05}.   

These spectacular results have been obtained despite the fact that most spectroscopic observations of roAp stars were
obtained in a snapshot mode and hence cover only a few hours of stellar oscillations. A detailed frequency analysis is highly
ambiguous with such limited data, because of the well known aliasing problem. It can only marginally improved by combining 
observations from several nights, but which rarely are available when large telescopes had to be used. 
Therefore, despite an outstanding clarity of the pulsation curves often produced with spectroscopy, 
for most multiperiodic roAp stars they correspond to an unresolved mixture of different modes and thus
are exceedingly difficult to interpret. This fundamental problem can be alleviated by simultaneous photometric
observations with a high duty cycle, as it was successfully demonstrated by a combination of space photometry of the roAp star HD\,24712
with contemporaneous VLT/UVES spectroscopy \citep{ryab2007a}. In this paper we
present an analysis of a much more extensive spectroscopic time series for another pulsating Ap star, 10\,Aql, which was observed
simultaneously from ground and space.

10\,Aql (HR\,7167, HD\,176232, HIP\,93179) is one of the brightest roAp stars. \citet{RSH00}
performed its model atmosphere and abundance analysis using high-resolution spectra. They have estimated
\teff\,=\,7550\,K, \lgg\,=\,4.0, and derived $M$\,=\,$2.0\pm0.2$\,$M_\odot$, $R$\,=\,$2.5\pm0.2$\,$R_\odot$ from the comparison of the
stellar parameters with theoretical evolutionary tracks. \citet{KB06} have included
10\,Aql in their study of the evolutionary state of magnetic chemically peculiar stars. Adopting a photometric
temperature \teff\,=\,7900\,K, they found $M$\,=\,$1.95\pm0.04$\,$M_\odot$, $\log L$\,=\,$1.32\pm0.05$ and a
fractional age of 64--76\% of the main sequence lifetime. The abundance pattern of 10\,Aql is characteristic of
a cool Ap star. A notable spectral anomaly is the overabundance of the doubly ionized REEs, \pr\ and \nd.

A longitudinal magnetic field of about 500\,G was detected in 10\,Aql by \citet{B58}, \citet{P70} and \citet{RWA05}. 
Using magnetic Zeeman splitting in unpolarized spectra \citet{KLR02} provided the first measurement
of the mean magnetic field modulus in 10\,Aql, \bs\,=\,$1.5\pm0.1$\,kG, which later was confirmed by
Leone, Vacca \& Stift (\citeyear{LVS03}),  and he estimated
the projected rotational velocity \vs\,=\,2.0\,\kms. The sharpness of spectral lines and the absence of measurable
variations of the mean longitudinal magnetic field indicate that 10\,Aql has a long rotational period.

Recently, Ryabchikova, Kochukhov \& Bagnulo (\citeyear{RKB08}) have studied the vertical stratification and isotope anomaly of Ca in a number
of Ap stars, including 10\,Aql. These authors found a concentration of Ca in deeper atmospheric layers and invoked
a vertical separation of the Ca isotopes to explain the shape of the Ca\ii\ infrared triplet lines. A
comprehensive chemical stratification analysis of 10\,Aql is currently underway by Nesvacil et al (in preparation), but already a preliminary
result (Nesvacil, Weiss \& Kochukhov \citeyear{N08}) shows that Si, Ca, Cr, Fe, and Sr share a qualitatively similar vertical
abundance distribution, characterized by a rapid increase of the element abundances below  $\log\tau_{5000}\approx
-1$. A re-determination of the atmospheric parameters taking into account individual chemical abundances and
stratification supports the spectroscopic \teff\,$\approx$\,7600\,K obtained by \citet{RSH00}.

Photometric variations with a
period close to 11.4\,min and an amplitude below 0.5\,mmag were discovered in 10\,Aql by  \citet{HK88}.  \citet{HK90}  identified three pulsation 
modes with periods of 11.06--13.45\,min, based on 26\,h of high-speed photometry in the $B$-band. Despite an aliasing problem in
their data, \citet{HK90} were able to determine the large frequency separation, $\Delta\nu=50.6$\,$\mu$Hz. 
Belmonte, Martinez Roger \& Roca Cortes (\citeyear{BMR91}) collected 47\,h of time-resolved photometry in the
$J$-band, confirming oscillations in 10\,Aql. Moreover, their study suggested that the amplitude of the
infrared photometric variations significantly exceeds variations in the $B$-band, which is unusual for a roAp
star (e.g. Martinez, Sekiguchi \& Hashimoto
\citeyear{MSH94}; \citealt{MWR96}). However, no oscillations were detected
in the follow-up photometric monitoring conducted in the $H$-band
(Belmonte, Kreidl \& Martinez Roger \citeyear{BKM92}). 

Radial velocity variations with previously known oscillation periods were discovered in 10\,Aql by \citet{KLR02}. Over two nights they observed 
the star during a total of 8\,h in a short wavelength region centred at
$\lambda$\,6150\,\AA. Unlike the majority of other roAp stars, 10\,Aql showed only a weak variability with an amplitude of
$\approx$\,30\,\ms\ in the strong Nd\iii\ 6145\,\AA\ line, but exhibited amplitudes at the level of 80--130\,\ms\
in the lines of first REE ions, Gd\ii\ and Eu\ii. Aiming to extend the study of pulsational variation
of 10\,Aql, \citet{HM05} analysed spectra covering 11.6\,h of time-resolved
observations during three nights. Somewhat surprisingly, despite a broad wavelength coverage of their data, \citet{HM05} 
were able to confirm the presence of RV variations in only 5 individual spectral lines,
two of which remained unidentified. The authors found the highest RV amplitude of 400\,\ms\ in an  unidentified line
at $\lambda$~5471.42\,\AA\ but could not detect pulsations in strong doubly ionized REE lines which show prominent
variability in other roAp stars. In addition to its remarkably low-amplitude photometric pulsational
variations, 10\,Aql seems to exhibit an unusual spectroscopic pulsational behaviour, making this star perhaps an intermediate object 
between high-amplitude roAp stars and apparently constant non-pulsating magnetic Ap stars.

10\,Aql was chosen as a target for an extensive observing campaign by the Canadian photometric space telescope MOST
\citep{W2003}. The star was observed as a primary target in June-July 2006, and nearly 120000
10\,s exposures were collected during 31\,d at a sampling interval of 20\,s. The frequency analysis by \citet{huber} 
revealed three definite periodicities and two candidate frequencies (see Table\,\ref{mostf}).
Huber et al. showed that the two largest
amplitude peaks reported by \citet{HK90} are 1\,d$^{-1}$ aliases of the intrinsic stellar
frequencies $f_1$ and $f_2$. The third significant MOST frequency, $f_3=1.43$\,mHz, has not been detected in
previous photometric observations of 10\,Aql. The lack of rotational modulation signal in the MOST data confirms the
suspected slow rotation of the star. \citet{huber} derived an improved value of the large frequency
separation, $\Delta\nu=50.95$\,$\mu$Hz, and have compared the observed frequencies with the predictions of the
theoretical models of non-adiabatic, magnetically distorted {\it p-}modes \citep{S05}.

\begin{table}
\caption{Pulsation frequencies of 10\,Aql identified in the MOST photometric campaign
\citep{huber}. The numbers in brackets give the estimated error in units of the last
significant digit. The two candidate frequencies, $f_i$ and $f_j$, are possibly present
in the MOST data but could not be definitely confirmed. 
\label{mostf}
}
\begin{tabular}{lllc}
\hline
id & $\nu$ (mHz) & $P$ (min) & $A$ (mmag) \\
\hline
$f_1$ & 1.44786(3) & 11.5112(2) & 0.17(1) \\
$f_2$ & 1.39691(3) & 11.9311(3) & 0.15(1) \\
$f_3$ & 1.42709(4) & 11.6788(3) & 0.12(1) \\
$f_i$ & 1.3662(1)  & 12.199(1)  & 0.03(2) \\
$f_j$ & 1.4686(1)  & 11.349(1)  & 0.03(2) \\
\hline
\end{tabular}
\end{table}

With the aim to improve the frequency analysis, study line profile variations and perform a tomographic analysis of the
pulsations in the atmosphere of 10\,Aql, we have organized a ground-based spectroscopic observing campaign
simultaneously with the MOST observations. Using high-resolution spectrographs at 4- and 8-m telescopes,  we have
collected a superb observational material with very high signal-to-noise ratio, wide wavelength
coverage and a high spectral resolving power. This combination of the data quality and time coverage has been never
achieved before for any roAp star. 

Preliminary results of our pulsation study of 10\,Aql were reported by \citet{SKR07}. In that paper
we outlined the frequency analysis procedure and discussed the main pulsation properties of
10\,Aql, such as the presence of a node-like behaviour in the bisector variations, phases and
amplitudes of radial velocity variation for different REE ions. Subsequently, Elkin, Kurtz \&
Mathys (\citeyear{EKM08}) published an analysis of a short spectroscopic UVES time series of
10\,Aql. These data, covering only 2 hours, were analysed without taking the multiperiodic character into account. 
The pulsation frequency of 1.428\,mHz adopted by Elkin et al. coincides with neither of the
two highest-amplitude pulsation modes in our data, but is close to the frequency $f_3$ identifyed by
\citet{huber} and by our spectroscopic analysis (see below).

The present paper is organized as follows. In Sect.\,\ref{observ} we describe spectroscopic observations and data reduction.
Section\,\ref{RV} describes methodology of the line identification and radial velocity measurements. The frequency
analysis is outlined in Sect.\,\ref{freq}. We establish a relation between photometric and spectroscopic variations of
10\,Aql in Sect.\,\ref{phase}. Line profile variations, indicating a pulsation
node in the stellar atmosphere, and the discovery of a remarkable asymmetry in the \nd\ lines are presented in
Sect.\,\ref{puls}. The paper ends with conclusions and discussion in Sect.\,\ref{discus}.

\section{Observations and data reduction}
\label{observ}

\subsection{Time-series of UVES spectra}

The main set of observational material analysed in this paper was obtained at the ESO 8-m VLT UT2 telescope using
the UVES spectrograph. 10\,Aql was observed for about 4\,h during each
of the following four nights in 2006: July 3, 9, 15, and 17. Each subset of observations consisted of 211 stellar
exposures, followed by a ThAr lamp spectrum. We have used UVES in the 600~nm red-arm setting, that provided access
to the 4980--6980\,\AA\ spectral window. The wavelength coverage is complete, except for a  $\approx$\,100~\AA\ gap
centred at $\lambda$~6000~\AA.  To facilitate the analysis of the narrow line profiles of 10\,Aql and to improve the
wavelength stability we used the UVES image slicer No.~3, resulting in a spectral resolving power of $\lambda/\Delta\lambda\approx 115\,000$.

10\,Aql was observed with exposure times of 40~s. The fast readout mode of the UVES CCDs (625~kpix\,s$^{-1}$,
4-port, low gain) reduces the overhead to about 29~s, resulting in a time resolution of 69~s. The typical
signal-to-noise ratio of individual spectra is 250--300 at $\lambda$\,=\,5000--5500~\AA.

Reduction of the UVES spectra utilized a set of echelle spectral processing routines developed by
Tsymbal, Lyashko, \& Weiss (\citeyear{tsymbal}) and
Lyashko, Tsymbal \& Makaganuik (\citeyear{lyashko}). Their code performs the standard reduction steps (averaging of
calibration frames, determination of the echelle order position, bias subtraction, flat fielding, extraction of 1-D
spectra and wavelength calibration) and is optimized to treat the UVES slicer spectra. Further details of the
application of this spectral reduction package to the time-resolved UVES observations of roAp stars are given by
\citet{ryab2007b}. This paper also details the final reduction step in which we improve
continuum normalization of the individual time-resolved spectra obtained in the same night. Observations
were obtained in different nights, hence we have carried out an additional continuum rectification correction
to achieve a global consistency of the continua in all 844 UVES spectra.

Our dataset of the UVES time-resolved spectra was complemented with the observations obtained
in the context of program  077.D-0150 \citep{EKM08} on July 24, which were taken with the instrumental setup
identical to ours and consist of 105 exposures obtained with the same exposure time and sampling
as in our observations. These spectra along with the appropriate calibrations were retrieved from
the ESO Archive and were subsequently processed in the same way as our own observations. Thus, a
total of 949 high-quality UVES spectra are available for the pulsational analysis of 10\,Aql.

\begin{table*}
\caption{Journal of spectroscopic observations of 10\,Aql.}
\label{tbl1}
\begin{tabular}{lcccccccl}
\hline
Date & Start HJD  & End HJD & Spectral range &N spec.&Exposure&  Time          & Typical & Instr. \\
     & (245 0000+)&(245 0000+)& (\AA)        &       &time (s)& resolution (s) & S/N &  \\
\hline
03/07/2006&3919.60523&3919.77510&4980--6980&211&40& 69&330 &UVES\\
09/07/2006&3925.64884&3925.81471&4980--6980&211&40& 69&260 &UVES\\
15/07/2006&3931.62381&3931.79193&4980--6980&211&40& 69&240 &UVES\\
17/07/2006&3933.63894&3933.80793&4980--6980&211&40& 69&300 &UVES\\
24/07/2006&3940.64279&3940.72682&4980--6980&105&40& 69&300 &UVES\\
14/07/2006&3930.52693&3930.61096&4572--7922&59 &70&120&150 &SARG\\
15/07/2006&3931.52970&3931.66690&4572--7922&86 &70&120& 90 &SARG\\
16/07/2006&3932.49220&3932.58039&4572--7922&61 &70&120&140 &SARG\\
\hline
\end{tabular}
\end{table*}

\subsection{Time-series of SARG spectra}

In addition to the UVES time series, 10\,Aql was observed at high
resolution with the spectrograph SARG at the
3.55-m Telescopio Nazionale Galileo (TNG) at the Observatorio del
Roque de los Muchachos (La Palma, Spain) in the
three consecutive nights of July 14--16, 2006. The total number of
echelle spectra obtained with the SARG is 206.
The star was monitored for 2.0--3.3~h using 70~s exposure times in
each of the three nights. Including the
50~s overhead, the time resolution of these observations is 120~s.
The SARG spectra cover the range of
4572--7922~\AA\ without gaps with a resolving power of about 86\,000.
The typical S/N ratio achieved for individual
exposures is 90--150 at $\lambda$\,=\,5000--5500~\AA.

The spectra were reduced with standard procedures for
spectroscopic observations which are part of the
Image Reduction and Analysis Facility package of NAOA 
in the same way as in Leone \& Catanzaro \cite{LC04}. The post-processing of 1-D
extracted SARG spectra was
performed consistently with the UVES spectra.

\section{Line identification and radial velocity measurements}
\label{RV}

To perform a careful line identification and to choose suitable lines for pulsation analyses, we have synthesised
the whole spectral region with the model atmosphere parameters \teff\,=\,7550\,K, \lgg\,=\,4.0, and abundances from
\citet{RSH00}. The mean magnetic field modulus of \bs\,=\,1.5\,kG \citep{KLR02} was adopted. Atomic data were 
extracted from the VALD database \citep{vald2}, the revised
atomic parameters for \nd\ were taken from \citet{RRKB06}, and the synthetic spectrum
calculations were carried out with the SYNTHMAG code (\citealt{P99}; \citealt{K06}).

We have measured radial velocities using the centre-of-gravity technique as outlined by \citet{KR01}. A relatively low 
pulsation amplitude and a small number of spectral lines showing pulsation
signatures seriously limited the usability of identification lists developed for other roAp
stars. Therefore, we started by measuring practically all spectral lines (about 2000) in the July 3 time
series in order to find lines with useful pulsation signatures. It turns out that the total number of such lines, both
blended and unblended, is about 150. Most of the other lines belong to iron-peak elements 
which do not show significant RV variations or are blended to an unacceptable degree.

It is a well-known fact that spectroscopic pulsational variability of typical roAp stars is dominated by the lines
of rare-earth ions, especially those of singly and doubly ionized Pr and Nd, which are strong and numerous in the
roAp spectra (see, for example, Savanov, Malanushenko \& Ryabchikova \citeyear{SMR99}; \citealt{KR01}). 10\,Aql is
different in this respect from other  roAp stars. REE lines  are weak, in particular those of the first ions. Their
equivalent widths are confined between 2 and 12~m\AA. The Pr and Nd overabundance inferred from the lines of the
first ions does not exceed 1.0~dex. Radial velocity measurements from these very weak lines often have insufficient
precision for the purpose of our investigation. Therefore, we used for our further analyses only the lines with an
equivalent width larger than 5~m\AA. The line identification list compiled for the July 3 spectra was then applied
to all other UVES observations, using consistent central wavelengths and measurement windows. For strong and
intermediate-strong lines our measurements of individual radial velocities have an internal accuracy of up to 10~\ms. 

We have unambiguously detected pulsations in the core of \ha\ and in the lines of Y\ii, La\ii,
Ce\ii, Pr\iii, Nd\ii, Nd\iii, Sm\ii, Eu\ii, Gd\ii, Tb\iii,
and Dy\iii. All lines show strong changes in RV amplitude, indicative of beating between several excited
modes. Typical observed semi-amplitudes, averaged over the whole duration of our spectroscopic monitoring, range
from 50 to 150~\ms. During the periods of constructive interference amplitudes approach 400~\ms\ for some REE lines.

Remarkably large RV amplitudes -- 350~\ms\ on average -- were measured for the \dy\ 5345,
5556, 5730~\AA\ lines and for the two unidentified lines at $\lambda$~5373.02 and 5471.40~\AA. The identification
of 5345 and 5556~\AA\ lines is based on the unpublished data kindly provided by A. Ryabtsev (Institute of Spectroscopy RAS).
All these lines have
equivalent widths in the range of 9--12~m\AA\ and their depths are less than 9\% of the continuum. The pulsational
characteristics of the two unidentified features are similar to those of \dy\ and a few Ce\ii\ lines. Without
any doubt both unidentified lines belong to the REE species. Both features are measured in
HD~24712 (\citealt{ryab2007a}) and their pulsational characteristics are much closer to \dy\ lines than to
Ce\ii\ lines. In $\gamma$~Equ, too, \dy\ 5730~\AA\  and 5373.02, 5471.40~\AA\ lines have similar
pulsational amplitudes and phases, which differ significantly from those of Ce\ii\ lines. Thus, we suggest
that the unidentified lines at 5373.02 and 5471.40~\AA 5471.40 belong to the \dy\ spectrum. 

The strong \nd\ and \pr\ lines in 10\,Aql, that usually exhibit the largest amplitudes in roAp stars, have
amplitudes below 100~\ms\ and probably form in the nodal area (see Sect.~\ref{puls}). The strongest Nd\ii\
lines in the observed spectral range have equivalent widths smaller than 7~m\AA, while Pr\ii\ lines are  not
measurable at all. Hence, the blending problem in this star becomes more serious.  Even slight blending of a REE
feature may significantly decrease the inferred amplitude \citep{KLR02}.
Taking into account the observed pulsation amplitudes and accuracy of RV measurements (defined primarily by
the spectral line strength and central wavelength), we have selected 65 lines for the frequency analysis and the
study of the pulsational wave propagation through the stellar atmosphere.
This line list is provided in Table~\ref{lines}.  

We also report on the analysis of a subset of non-pulsating lines belonging to the iron-peak elements, Ca, and Ba.
These results, given in Table~\ref{nplines}, demonstrate the overall accuracy of our pulsation measurements.  Among the
lines with definite radial velocity variation the lowest amplitude of 15$\pm$1~\ms\ is detected in the  strong
Y\ii~5662.9 \AA\ line ($W_\lambda$\,=\,57~m\AA). In comparison, for the non-pulsating lines of the same intensity the
amplitudes are at the level of 2--4~\ms. We consider this to be the upper limit of pulsation amplitude for the lower
atmosphere of 10\,Aql. Note that although for many non-pulsating lines the free-period search yields values comparable
to real frequencies of 10\,Aql (see next Section), the False Alarm Probability \citep{horne} of such signals is always
large (Prob, listed in the Table is small).

In addition to the centre-of-gravity RV measurements, we studied bisector velocities across the H$\alpha$
line core and profiles of the strongest REE spectral lines.

The UVES and SARG datasets overlap for about 1~h in the night of July 15. The radial velocities from the two
spectrographs agree within the error bars, where the RV measurements with the SARG spectra are less accurate
due to a lower S/N.

\section{Frequency analysis}
\label{freq}

We carried out the frequency analysis by applying the standard combination of discrete Fourier transform (DFT) and
least-squares fitting. For each of the studied REE ions we found a number of lines showing clear pulsation signatures.
Aiming to improve the signal-to-noise ratio of the RV data, we subtracted linear trends from the measurements of different
lines of the same ions and averaged these data. The RVs derived from the average of 5
\dy\ lines turned out to be especially useful for the frequency analysis. In all nights of our UVES observations, 10\,Aql
exhibited a large variation in the RV amplitude, which often changed from $<$\,100~\ms\ to 500--700~\ms\ in the course of
a few hours. This behaviour, illustrated in Fig.~\ref{dy3_rv} for \dy, is caused by beating of several frequencies.

\begin{figure}
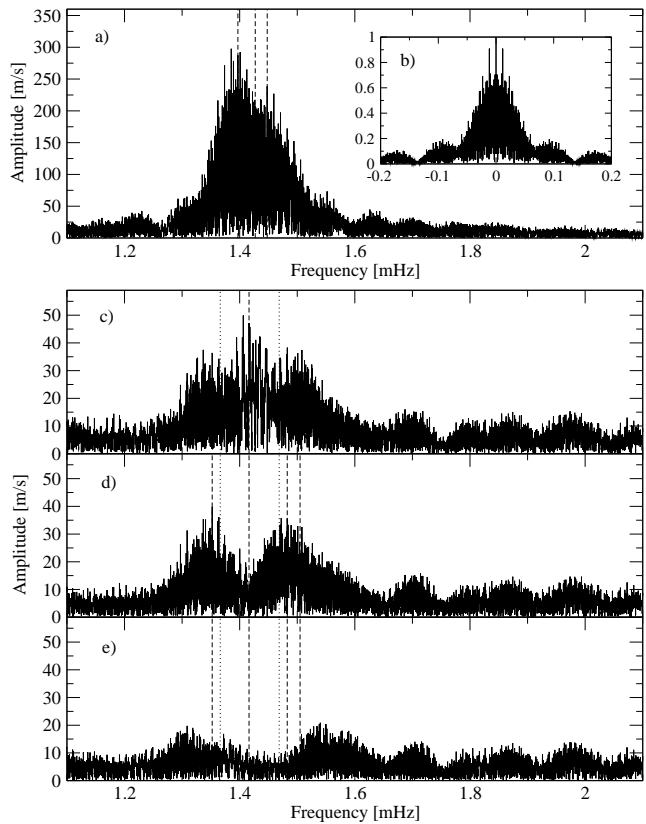

\figps{10_Aql_fig2.eps}
\caption{The amplitude spectra for the average \dy\ radial velocities inferred from the UVES+SARG observations.
\textit{a)} the DFT of the radial velocity signal; \textit{b)} spectral window;
\textit{c)} the residual amplitude spectrum
obtained after prewhitening the three main frequencies $f_1$--$f_3$ (dashed lines);
\textit{d)} the residual amplitude spectrum
obtained after prewhitening the four frequencies $f_1$--$f_4$;
\textit{e)} the residual noise after prewhitening the seven frequencies $f_1$--$f_7$ from Table~\ref{specf}. These
frequencies are indicated by dashed lines, while the dotted lines correspond to the MOST candidate frequencies.}
\label{dy3_ft}
\end{figure}

\begin{figure}
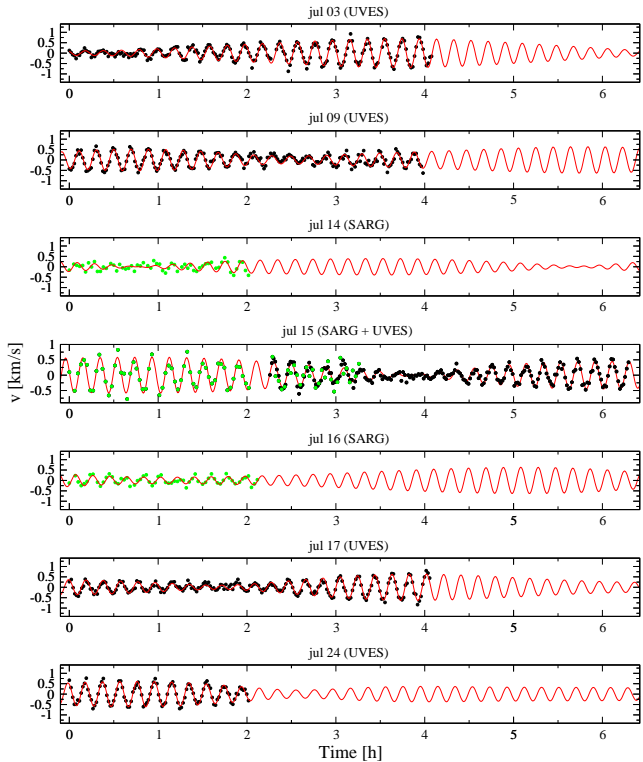

\figps{lcs_new.eps}
\caption{The average radial velocity variation of \dy\ lines during seven nights of spectroscopic observations. The measurements
are shown with black (UVES) and green (SARG) dots. The solid curve represents a least-squares fit of seven pulsation 
frequencies detected in 10\,Aql.}
\label{dy3_rv}
\end{figure}

The total time span of the spectroscopic data set is about 25 hours, but the spectra were obtained in a period of 21
days and, thus, are characterized by a very low duty cycle. Despite the clarity of the pulsational RV variations, severe
aliasing inhibits a secure identification of the frequencies from the spectroscopic data alone. For example, the peak
with a highest amplitude is often found at 1.3853~mHz, which is a d$^{-1}$ alias of the photometric frequency $f_2$.
Figure~\ref{dy3_ft} illustrates this problem for \dy.

The original motivation for simultaneous MOST and spectroscopic observations was to take advantage of the accurate
photometric frequency information in the analysis of spectroscopic data. Therefore, we have adopted the three main MOST
frequencies as the initial guess in a non-linear least-squares fitting procedure. Then, we optimized the amplitudes,
phases and periods of the three frequency components for each REE ion. 
For a number of REE ions (Ce\ii, \nd, Sm\ii, Eu\ii,
Gd\ii, Tb\iii, \dy) we were able to determine all three frequencies with a S/N\,=\,10--60. The formal
accuracy of the periods determined for each of these ions is comparable or better than for the MOST  photometry. We
attribute this result to a much higher S/N achieved in the spectroscopic observations. In Table~\ref{specf}
we report the weighted mean of the spectroscopic pulsation frequencies in 10\,Aql. A comparison with MOST
(Table~\ref{mostf}) shows that all our frequencies deviate by less than 1~$\sigma$ from the photometric results.

\begin{figure}
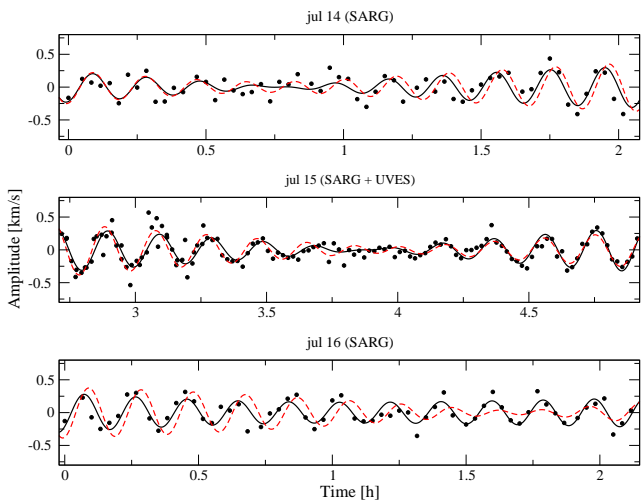

\figps{10_Aql_fig3.eps}
\caption{Part of the average Dy\iii\ radial velocity curve (dots) around the amplitude minimum.
The three-frequency and seven-frequency solutions are shown by dashed and solid lines,
respectively.}
\label{beat}
\end{figure}

\begin{table}
\caption{Secure (bold) and tentative (italics) spectroscopic pulsation frequencies of 10\,Aql determined from the analysis of the average RVs of 7 REE ions derived
from the UVES spectra (second and third columns) and from the average RVs of 5 Dy\iii\ lines measured in the UVES+SARG spectra (fourth and fifth columns).
The numbers in brackets give the error in units of the last significant digit.
\label{specf}
}
\begin{center}
\begin{tabular}{l|cc|cc}
\hline
id & $\nu$ (mHz)   & ion  & $\nu$ (mHz) & S/N \\
\hline
&\multicolumn{2}{c|}{UVES data}&\multicolumn{2}{c}{UVES+SARG data, Dy\iii}\\
\hline
$f_1$ &{\bf 1.447842(4)} & all &{\bf 1.44787(1)}&39.8\\
$f_2$ &{\bf 1.396883(3)} & all &{\bf 1.39684(1)}&48.2\\
$f_3$ &{\bf 1.427082(7)} & all &{\bf 1.42711(1)}&21.2\\
\cline{2-5}
$f_4$ &{\it 1.4108}&Dy\iii , Tb\iii&{\it 1.41627(3)}&7.9\\
      &{\it 1.4341}&Ce\ii &&\\
      &{\it 1.4177}&Sm\ii &&\\
$f_5$ &{\it 1.3127}&Dy\iii &{\it 1.35227(3)}&6.6\\
      &{\it 1.3367}&Ce\ii &&\\
      &{\it 1.3424}&Sm\ii &&\\
      &{\it 1.3600}&Nd\iii &&\\
$f_6$ &            &             &{\it 1.48260(3)}&6.0\\
$f_7$ &{\it 1.5048}&Dy\iii &{\it 1.50488(5)}&4.3\\
      &{\it 1.5125}&Sm\ii &&\\
      &{\it 1.5252}&Gd\ii &&\\

\hline
\end{tabular}
\end{center}
\end{table}

After prewhitening the radial velocity curves with the three main frequencies, all REEs show residual amplitude peaks in
the 1.3--1.5~mHz range. In the UVES data the residual RV
variations have amplitudes between 10 and 50~\ms, and show different frequencies at the S/N level of 3--4, depending on
the ion. None of these frequencies coincide with the candidate MOST frequencies $f_i$ and $f_j$. Accurate determination
of the low-amplitude frequencies in spectroscopy is hampered by the aliasing problem. Nevertheless, we note that the
most promising spectroscopic candidate frequency is $f=1.3367$~mHz. Variation with this period (or its aliases) is seen
in \dy, Ce\ii, Sm\ii, and \nd. This frequency coincides with a sidelobe of a MOST orbital
frequency harmonics and therefore does not contradict the space photometry reported by \citet{huber}, because these 
frequency ranges were excluded deliberately from the analysis. The other
candidate spectroscopic frequencies are 1.411--1.434~mHz (seen in \dy, Tb\iii , Ce\ii , and Sm\ii)
and 1.505--1.525~mHz (\dy, Sm\ii, and Gd\ii). They are not observed in the MOST data.
The amplitudes and phases of the three
main periods do not change beyond their formal error bars when additional low-amplitude frequencies are included in the
fit. 

A lower S/N and the specific temporal coverage of the SARG observations close to minimum RV due to beating effects
do not allow us to use these RV measurements for all REE ions. However, we found that the highest amplitude \dy\ lines
allow to analyse the complete set of UVES+SARG data and improve the frequency solution.

Three practically unblended Dy\iii\ lines (5345.37, 5556.13, and 5730.34 \AA) as well as two unidentified lines
(5373.02 and 5471.42~\AA) that we propose to identify also as Dy\iii\ were chosen for RV measurements. The linear
instrumental trends were
subtracted for each night of observations. Because of the negligible phase shift between the RV curves of these lines we can 
average the
RV's and weight data according to the individual errors to increase the accuracy of the frequency analysis. 
After subtracting the three main frequencies, that are the same as detected by MOST or by using only the UVES data,
the residual RV variations
show amplitudes clearly exceeding the noise level in three distinct regions, indicating the presence of additional signal 
in the data (see Fig.\,\ref{dy3_ft}c).
To reduce the aliasing problem the program
{\sc SigSpec} \citep{sigspec} was applied to the residuals for an accurate determination of additional low-amplitude frequencies.
The main functionality of {\sc SigSpec} is the evaluation of the probability that a signal in the Fourier-domain is not 
caused by random noise by taking frequency, amplitude and phase information into account, yielding a quantity called spectral significance.
In addition, {\sc SigSpec} employs an anti-aliasing correction procedure, in which several trial frequencies 
(starting with the one showing the highest significance) are prewhitened. The most probable frequency is assumed to be that which results in the lowest 
residual scatter of the data after prewhitening.
The program Period04 \citep{period04} was used to check the {\sc SigSpec} solution by calculating a multi-sine fit, allowing simultaneously to improve frequency,
amplitude and phase of all detected signals.

\begin{table*}
\caption{Summary of our spectroscopic pulsational analysis of individual spectral lines in 10\,Aql. The columns give
central wavelengths in \AA, followed by the free search period $P$ with the respective error estimate $\sigma_P$ derived with a least-squares fit.
The corresponding probability of a periodic signal (Prob.) is calculated according to \citet{horne}.
The next three groups of columns give pulsation amplitudes and phases $\varphi$  
with the respective errors,
determined with a simultaneous fit of three fixed pulsation periods indicated in the column head. Phases 
are given 
as a fractional pulsation period.
The starting point for frequency calculations is HJD=24453919.60523.
\label{lines} }
\begin{tabular}{c|ccc|rrcc|rrcc|rrcc|r}
\hline
 &\multicolumn{3}{c|}{Free period} & \multicolumn{12}{c|}{Fixed periods }& \\
$\lambda$      &         & & &\multicolumn{4}{c|}{$P$\,=\,11.931 min} &\multicolumn{4}{c|}{$P$\,=\,11.511 min} &\multicolumn{4}{c|}{$P$\,=\,11.679 min} & $W_\lambda$ \\
        (\AA)   & $P$ (min) & $\sigma_P$ & Prob.& $A$ &$\sigma_A$ & $\varphi$ & $\sigma_{\varphi}$ & $A$ &$\sigma_A$ &$\varphi$&$\sigma_{\varphi}$& $A$ &$\sigma_A$ &$\varphi$&$\sigma_{\varphi}$&(m\AA)\\
\hline
\multicolumn{17}{l}{H$\alpha$ core} \\
 6562.799 &11.42103 & 0.00010 & 1.0000 &  52 &  2 & 0.611 & 0.007 & 49 &  2 & 0.287 & 0.007 & 37 &  2 & 0.826 & 0.010 &  \\[1mm]
\multicolumn{17}{l}{Y\ii} \\
 5662.934 &12.03117 &0.00015  & 1.0000 &  15 &  1 & 1.100 & 0.013 &  5 &  1 & 0.797 & 0.035 &  4 &  1 & 1.251 & 0.026 &57.3\\[1mm]
\multicolumn{17}{l}{La\ii} \\
 5290.814 &11.41958 &0.00037  & 0.9996 &  18 & 12 & 0.645 & 0.111 & 61 & 12 & 1.174 & 0.033 & 23 & 13 & 0.715 & 0.092 & 4.0\\
 5377.074 &11.68604 &0.00032  & 1.0000 &  27 & 13 & 0.434 & 0.078 & 78 & 13 & 1.195 & 0.027 & 81 & 14 & 0.742 & 0.028 & 5.2\\
 5805.773 &11.51219 &0.00023  & 1.0000 &   1 &  5 & 0.946 & 0.132 & 44 &  5 & 1.313 & 0.018 & 29 &  5 & 0.838 & 0.029 & 8.9\\[1mm]
\multicolumn{17}{l}{Ce\ii} \\
 5037.803 &12.03132 &0.00013  & 1.0000 &  77 &  7 & 0.861 & 0.015 & 61 &  7 & 0.480 & 0.019 & 74 &  7 & 0.986 & 0.017 & 6.3\\
 5044.018 &12.03110 &0.00007  & 1.0000 & 129 &  5 & 0.856 & 0.006 &109 &  5 & 0.502 & 0.008 & 90 &  5 & 0.998 & 0.010 & 7.1\\
 5274.231 &12.03107 &0.00005  & 1.0000 & 263 &  5 & 0.888 & 0.003 &186 &  5 & 0.555 & 0.005 &141 &  6 & 1.055 & 0.007 &12.7\\
 5330.548 &12.03110 &0.00009  & 1.0000 & 116 &  5 & 0.885 & 0.008 & 92 &  5 & 0.514 & 0.010 & 76 &  6 & 1.073 & 0.013 & 6.6\\
 5468.391 &12.03106 &0.00009  & 1.0000 & 189 &  7 & 0.893 & 0.007 &131 &  7 & 0.524 & 0.009 &100 &  8 & 1.043 & 0.013 & 7.5\\
 5512.047 &12.03094 &0.00008  & 1.0000 & 121 &  4 & 0.857 & 0.005 & 87 &  4 & 0.501 & 0.008 & 61 &  4 & 1.021 & 0.011 &10.0\\
 5695.846 &12.03110 &0.00020  & 1.0000 & 182 & 20 & 0.824 & 0.018 &132 & 20 & 0.444 & 0.025 &111 & 21 & 0.991 & 0.031 & 3.2\\[1mm]
\multicolumn{17}{l}{\pr} \\
 5299.988 &12.03111 &0.00021  & 1.0000 &  25 &  3 & 1.318 & 0.020 & 16 &  3 & 0.923 & 0.031 &  4 &  3 & 1.500 & 0.128 &38.5\\
 6090.013 &11.33115 &0.00033  & 0.9945 &  13 &  3 & 1.352 & 0.042 & 18 &  3 & 0.903 & 0.031 &  9 &  3 & 1.441 & 0.062 &19.1\\
 6160.248 &11.42069 &0.00029  & 1.0000 &  25 &  3 & 1.290 & 0.022 & 24 &  3 & 0.812 & 0.023 &  6 &  3 & 1.469 & 0.087 &19.8\\
 6195.596 &12.04860 &0.00044  & 0.9991 &  20 &  5 & 1.243 & 0.042 &  5 &  5 & 0.804 & 0.145 &  8 &  5 & 1.593 & 0.105 &35.2\\[1mm]
\multicolumn{17}{l}{Nd\ii} \\
 5130.593 &11.89818 &0.00011  & 1.0000 &  67 &  6 & 0.898 & 0.014 & 71 &  6 & 0.527 & 0.014 & 59 &  6 & 1.012 & 0.018 & 6.3\\
 5293.156 &11.48186 &0.00019  & 1.0000 &  34 &  4 & 0.845 & 0.019 & 25 &  4 & 0.524 & 0.027 & 14 &  4 & 1.017 & 0.052 & 6.1\\
 5319.821 &11.89797 &0.00014  & 1.0000 &  85 &  8 & 0.865 & 0.016 & 88 &  8 & 0.497 & 0.016 & 62 &  9 & 1.006 & 0.024 & 5.7\\[1mm]
\multicolumn{17}{l}{\nd} \\
 5102.435 &12.03096 & 0.00010 & 1.0000 &  35 &  1 & 1.215 & 0.009 & 10 &  1 & 0.862 & 0.029 &  2 &  2 & 1.484 & 0.126 &110.~\\
 5286.742 &12.03119 & 0.00014 & 1.0000 &  88 &  7 & 0.913 & 0.014 & 82 &  7 & 0.588 & 0.015 & 54 &  8 & 1.068 & 0.025 & 8.6\\
 5294.091 &11.93130 & 0.00015 & 1.0000 &  23 &  2 & 1.197 & 0.014 &  9 &  2 & 0.826 & 0.037 &  5 &  2 & 1.353 & 0.061 &114.~\\
 5633.560 &12.03131 & 0.00015 & 1.0000 & 126 &  9 & 0.952 & 0.012 & 82 &  9 & 0.625 & 0.018 & 70 &  9 & 1.105 & 0.022 &11.5\\
 5677.167 &12.03097 & 0.00016 & 1.0000 &  91 &  7 & 1.011 & 0.013 & 56 &  7 & 0.650 & 0.021 & 36 &  7 & 1.197 & 0.032 &23.3\\
 5802.534 &12.03121 & 0.00010 & 1.0000 &  92 &  4 & 0.983 & 0.007 & 67 &  4 & 0.632 & 0.010 & 43 &  4 & 1.128 & 0.016 &26.2\\
 5845.020 &12.03085 & 0.00014 & 1.0000 &  83 &  5 & 1.001 & 0.010 & 52 &  5 & 0.682 & 0.016 & 21 &  5 & 1.205 & 0.039 &35.6\\
 5851.531 &12.03104 & 0.00010 & 1.0000 &  93 &  4 & 0.965 & 0.007 & 60 &  3 & 0.645 & 0.010 & 37 &  4 & 1.175 & 0.017 &31.5\\
 6145.042 &11.93123 & 0.00026 & 1.0000 &  42 &  4 & 1.049 & 0.018 & 20 &  4 & 0.757 & 0.038 & 16 &  4 & 1.210 & 0.049 &90.5\\
 6327.260 &12.03120 & 0.00009 & 1.0000 &  54 &  1 & 1.001 & 0.005 & 36 &  1 & 0.676 & 0.008 & 19 &  1 & 1.175 & 0.015 &54.8\\
 6550.239 &12.03126 & 0.00011 & 1.0000 &  24 &  1 & 1.039 & 0.008 & 13 &  1 & 0.721 & 0.013 &  7 &  1 & 1.179 & 0.025 &58.2\\
 6690.765 &11.42080 & 0.00028 & 1.0000 & 105 & 17 & 0.985 & 0.026 &118 & 16 & 0.608 & 0.023 & 44 & 17 & 1.157 & 0.061 & 9.6\\[1mm]
\multicolumn{17}{l }{Sm\ii} \\
 5052.757 &12.03104 & 0.00007 & 1.0000 & 172 &  5 & 0.682 & 0.005 &189 &  5 & 0.333 & 0.004 &149 &  5 & 0.837 & 0.006 &14.1\\
 5069.441 &11.80060 & 0.00007 & 1.0000 & 151 &  6 & 0.683 & 0.007 &156 &  6 & 0.329 & 0.007 &132 &  7 & 0.844 & 0.009 &15.1\\
 5103.082 &12.03093 & 0.00007 & 1.0000 & 145 &  3 & 0.692 & 0.004 &146 &  3 & 0.323 & 0.004 &107 &  3 & 0.836 & 0.006 &16.5\\
 5116.686 &12.03100 & 0.00008 & 1.0000 & 184 &  8 & 0.670 & 0.007 &193 &  8 & 0.326 & 0.007 &141 &  8 & 0.836 & 0.010 & 9.4\\
 5759.515 &11.42085 & 0.00012 & 1.0000 & 106 &  8 & 0.688 & 0.012 &134 &  8 & 0.299 & 0.009 & 78 &  8 & 0.844 & 0.017 & 7.8\\[1mm]
\multicolumn{17}{l}{Eu\ii} \\
 5818.739 &11.79987 & 0.00025 & 1.0000 &  71 & 12 & 0.608 & 0.027 & 64 & 11 & 0.176 & 0.030 & 31 & 12 & 0.786 & 0.063 & 5.4\\
 6173.043 &11.89795 & 0.00025 & 1.0000 &  63 &  7 & 0.594 & 0.019 & 55 &  7 & 0.216 & 0.022 & 28 &  7 & 0.767 & 0.044 & 9.3\\
 6303.408 &12.03093 & 0.00020 & 1.0000 &  81 &  7 & 0.579 & 0.015 & 37 &  7 & 0.179 & 0.034 & 27 &  8 & 0.759 & 0.046 & 7.3\\
 6437.670 &12.03094 & 0.00016 & 1.0000 &  72 &  5 & 0.588 & 0.011 & 48 &  5 & 0.226 & 0.017 & 32 &  5 & 0.818 & 0.025 &26.1\\
 6645.070 &12.03091 & 0.00009 & 1.0000 & 108 &  3 & 0.624 & 0.006 & 70 &  3 & 0.254 & 0.009 & 52 &  3 & 0.834 & 0.012 &34.5\\
\hline
\end{tabular}
\end{table*}

\begin{table*}
\caption{Continuation of Tab.\,\ref{lines}. \label{lines-c}}
\begin{tabular}{c|ccc|rrcc|rrcc|rrcc|r}
\hline
 &\multicolumn{3}{c|}{Free period} & \multicolumn{12}{c|}{Fixed periods }& \\
$\lambda$      &         & & &\multicolumn{4}{c|}{$P$\,=\,11.931 min} &\multicolumn{4}{c|}{$P$\,=\,11.511 min} &\multicolumn{4}{c|}{$P$\,=\,11.679 min} & $W_\lambda$ \\
        (\AA)   & $P$ (min) & $\sigma_P$ & Prob.& $A$ &$\sigma_A$ & $\varphi$ & $\sigma_{\varphi}$ & $A$ &$\sigma_A$ &$\varphi$&$\sigma_{\varphi}$& $A$ &$\sigma_A$ &$\varphi$&$\sigma_{\varphi}$&(m\AA)\\
\hline
\multicolumn{17}{l}{Gd\ii} \\
 5092.246 &11.78693 & 0.00008 & 1.0000 & 100 &  6 & 0.807 & 0.010 &101 &  6 & 0.422 & 0.010 & 98 &  6 & 0.920 & 0.011 &12.2\\
 5372.208 &11.78713 & 0.00021 & 1.0000 &  28 &  8 & 0.719 & 0.049 & 66 &  8 & 0.204 & 0.021 & 68 &  9 & 0.831 & 0.023 & 5.6\\
 5419.876 &11.51224 & 0.00020 & 1.0000 &  12 &  6 & 0.567 & 0.083 & 67 &  6 & 0.194 & 0.016 & 54 &  7 & 0.775 & 0.021 & 5.4\\
 5469.707 &11.64521 & 0.00026 & 1.0000 &  29 & 12 & 0.691 & 0.067 & 88 & 12 & 0.217 & 0.023 & 82 & 13 & 0.819 & 0.025 & 5.0\\
 5500.432 &11.78666 & 0.00022 & 1.0000 &  57 & 14 & 0.750 & 0.040 & 99 & 14 & 0.205 & 0.023 & 87 & 15 & 0.809 & 0.028 & 4.4\\
 5583.669 &11.78669 & 0.00028 & 1.0000 &  69 & 12 & 0.754 & 0.028 & 57 & 12 & 0.302 & 0.033 & 45 & 12 & 0.847 & 0.044 & 5.7\\
 5721.971 &11.69107 & 0.00024 & 1.0000 &  22 & 11 & 0.859 & 0.082 & 99 & 11 & 0.172 & 0.018 & 79 & 11 & 0.780 & 0.024 & 4.7\\
 5733.856 &11.69121 & 0.00021 & 1.0000 & 121 & 14 & 0.841 & 0.019 & 92 & 14 & 0.434 & 0.024 & 99 & 14 & 0.947 & 0.023 & 9.9\\
 5815.839 &11.69088 & 0.00022 & 1.0000 &  17 &  6 & 0.786 & 0.061 & 60 &  6 & 0.189 & 0.018 & 46 &  6 & 0.764 & 0.024 & 8.8\\
 5840.468 &11.51229 & 0.00028 & 1.0000 &  15 & 13 & 0.725 & 0.144 & 98 & 13 & 0.212 & 0.022 & 69 & 13 & 0.761 & 0.032 & 5.1\\
 5856.964 &11.42080 & 0.00020 & 1.0000 &  13 &  8 & 0.815 & 0.096 & 84 &  8 & 0.169 & 0.016 & 59 &  8 & 0.714 & 0.023 & 6.4\\
 5860.725 &11.69118 & 0.00024 & 1.0000 &  56 & 13 & 0.799 & 0.039 & 66 & 13 & 0.226 & 0.033 & 87 & 14 & 0.771 & 0.026 & 7.8\\
 5877.231 &11.78691 & 0.00018 & 1.0000 &  34 &  7 & 0.723 & 0.035 & 70 &  7 & 0.219 & 0.017 & 70 &  7 & 0.800 & 0.018 & 6.3\\
 5913.517 &11.78672 & 0.00020 & 1.0000 &  51 &  7 & 0.771 & 0.022 & 64 &  7 & 0.257 & 0.018 & 52 &  7 & 0.863 & 0.022 & 9.1\\[1mm]
\multicolumn{17}{l}{Tb\iii} \\
 5505.370 &12.03100 & 0.00006 & 1.0000 & 238 &  6 & 1.339 & 0.004 &156 &  6 & 1.024 & 0.007 &117 &  6 & 1.510 & 0.009 &15.0\\
 5847.221 &12.03102 & 0.00013 & 1.0000 & 355 & 19 & 1.346 & 0.009 &257 & 19 & 1.042 & 0.012 &163 & 20 & 1.569 & 0.020 &10.5\\
 6092.884 &11.80040 & 0.00014 & 1.0000 & 115 &  7 & 1.344 & 0.010 &114 &  7 & 0.990 & 0.010 & 75 &  7 & 1.508 & 0.015 &10.5\\
 6687.648 &12.13203 & 0.00046 & 0.8814 & 133 & 32 & 1.386 & 0.039 &130 & 32 & 1.006 & 0.039 & 84 & 32 & 1.501 & 0.062 & 8.2\\[1mm]
\multicolumn{17}{l}{\dy} \\
 5345.374 &12.03102 & 0.00006 & 1.0000 & 228 &  5 & 0.850 & 0.004 &217 &  6 & 0.513 & 0.004 &164 &  6 & 1.008 & 0.006 & 8.9\\
 5373.015 &12.03108 & 0.00006 & 1.0000 & 315 &  9 & 0.857 & 0.005 &268 &  9 & 0.524 & 0.006 &212 & 10 & 1.030 & 0.008 & 9.0\\
 5471.416 &12.03106 & 0.00009 & 1.0000 & 341 & 13 & 0.888 & 0.006 &286 & 13 & 0.523 & 0.008 &196 & 14 & 1.024 & 0.012 & 9.0\\
 5556.132 &12.03097 & 0.00008 & 1.0000 & 349 & 11 & 0.875 & 0.005 &311 & 11 & 0.537 & 0.006 &208 & 11 & 1.045 & 0.009 & 7.8\\
 5730.335 &12.03104 & 0.00009 & 1.0000 & 319 &  8 & 0.889 & 0.004 &273 &  8 & 0.540 & 0.005 &200 &  9 & 1.051 & 0.007 &12.0\\[1mm]
\multicolumn{17}{l}{Ho\iii} \\
 5543.732 &12.03096 & 0.00010 & 1.0000 &  77 &  3 & 1.098 & 0.007 & 63 &  3 & 0.750 & 0.008 & 39 &  3 & 1.278 & 0.014 &11.3\\
\hline
\end{tabular}
\end{table*}

\begin{table}
\caption{Summary of the frequency analysis of non-pulsating lines in 10\,Aql. The columns give
central wavelengths in \AA, followed by the free search period $P$ with the respective error estimate
$\sigma_P$ derived with a least-squares fit. The corresponding probability of the periodic signal
(Prob.), is calculated according to \citet{horne}.
The last three columns give an estimate of the RV amplitude and error, and equivalent width of the line.
}
\label{nplines}
\begin{tabular}{c|ccclc|l}
\hline
 $\lambda$ & $P$  & $\sigma_P$ & Prob.& $A$   &$\sigma_A$ & $W_\lambda$ \\
 (\AA)     & \multicolumn{2}{c}{(min)}&      & \multicolumn{2}{c}{(\ms)} & (m\AA)      \\
\hline
\multicolumn{7}{l}{Ca\i} \\
6162.176 & 11.53066 & 0.00061 & 0.0015  &  6  &  2  & 127 \\
\multicolumn{7}{l}{Ca\ii} \\
5019.981 & 14.34773 & 0.00075 & 0.0000  &  4  &  1  & 112 \\
5021.147 & 14.27341 & 0.00072 & 0.1659  &  7  &  2  &  47 \\
\multicolumn{7}{l}{Ti\i} \\
5036.454 & 13.59938 & 0.00068 & 0.6997  & 12  &  3  &  10 \\
5038.397 & 14.34091 & 0.00081 & 0.5762  & 10  &  3  &   9 \\
\multicolumn{7}{l}{Ti\ii} \\
5336.793 & 11.57725 & 0.00052 & 0.1716  &  5  &  2  &  55 \\
5418.776 & 12.31386 & 0.00072 & 0.0036  &  6  &  2  &  35 \\
\multicolumn{7}{l}{Cr\i} \\
5296.691 & 11.72380 & 0.00049 & 0.0079  &  5  &  1  &  59 \\
5297.371 & 12.97623 & 0.00097 & 0.0000  &  3  &  1  &  93 \\
5300.744 & 14.27558 & 0.00084 & 0.5068  &  5  &  2  &  33 \\
5312.854 & 11.56547 & 0.00042 & 0.1617  &  8  &  2  &  26 \\
5318.385 & 11.78757 & 0.00037 & 0.6640  & 7    & 1  &  36 \\
5348.325 & 11.84174 & 0.00053 & 0.0000  &  6  &  2  &  67 \\
\multicolumn{7}{l}{Cr\ii} \\
5305.869 & 11.95541 & 0.00056 & 0.0013  &  4  &  1  &  71 \\
5310.692 & 11.69145 & 0.00050 & 0.0019  &  5  &  1  &  55 \\
5313.583 & 11.68307 & 0.00065 & 0.0000  &  3  &  1  &  75 \\
\multicolumn{7}{l}{Fe\i} \\
5410.912 & 12.19297 & 0.00068 & 0.0001  &  5  &  1  &  84 \\
5415.204 & 12.08059 & 0.00064 & 0.0102  &  4  &  1  & 111 \\
\multicolumn{7}{l}{Fe\ii} \\
5414.075 & 12.11682 & 0.00058 & 0.0007  &  5  &  1  &  42 \\
\multicolumn{7}{l}{Co\i} \\
5331.452 & 11.57328 & 0.00061 & 0.0238  & 12  &  4  &  14 \\
5342.701 & 11.88451 & 0.00058 & 0.0000  &  5  &  2  &  30 \\
5347.496 & 11.59717 & 0.00059 & 0.0165  &  9  &  3  &  12 \\
\multicolumn{7}{l}{Ni\i} \\
5035.359 & 11.95589 & 0.00046 & 0.2132  &  7  &  2  &  21 \\
\multicolumn{7}{l}{Ba\ii} \\
5853.684 & 11.85784 & 0.00064 & 0.0000  & 16  &  5  &  32 \\
\hline
\end{tabular}
\end{table}

For the residual RV variations, the selection of the frequency at $f=1.4163$\,mHz yields the lowest residual scatter. 
Note that this peak does not coincide with the
highest peak in the amplitude spectrum (see Fig.~\ref{dy3_ft}c). Choosing another (e.g., the highest amplitude) frequency
in this prewhitening step will yield an almost completely different set of subsequent frequency values, but not reaching the 
same low noise level.

After prewhitening of this signal two additional power excess regions remain (Fig.~\ref{dy3_ft}d).
In total, three additional frequencies are needed to obtain residuals that show only noise (Fig.~\ref{dy3_ft}e).
Table~\ref{specf} and \ref{lines-c} lists all derived frequencies.  

Figure~\ref{dy3_rv} shows a direct comparison of the RV observations and a seven-frequency solution, while Fig.\,\ref{beat} displays an amplified part
of the observations near beating. Clearly, using only the three definite MOST frequencies is
not sufficient to fully reconstruct the observed variations. While the
7 frequency fit is an improvement one has to stress that the frequency
identification is insecure due to the severe aliasing effects and it 
strongly depends - as was already mentioned - on the prewhitening
sequence.

The discrepancy that none of the spectroscopically identified low-amplitude frequencies show up in the MOST photometry can 
be explained by their expected amplitude in the photometry of about 30\,ppm, which is about 3 times the noise level as stated by \citet{huber}. 
This estimate comes from scaling the RV amplitude
of $f_{1}$ to the highest peak in the residual RV variations and assuming the amplitude ratios to be similar in photometry and 
spectroscopy.
Therefore, this additional signal might be lost in the noise.
On the other hand, it has also been suggested that
spectroscopic and photometric observations, probing different layers of the stellar atmosphere,
can yield different pulsation frequencies
(Kurtz, Elkin \& Mathys \citeyear{KEM06}). The clear
lack of signal around 1.41\,mHz in photometry, but which shows a convincingly high S/N in spectroscopy,
could support this concept. That neither of the photometric candidate frequencies $f_{\rm i}$ nor $f_{\rm j}$
contribute to a good frequency solution for the RV variations corroborate that these frequencies are insecure, as was stated by \citet{huber}, or indicate that
the amplitude scaling between photometry and spectroscopy is not as simple as we assumed. The spectroscopic data does not show any signal close to the
third frequency ($\sim 1.24\,$mHz) proposed by \citet{HK90} and \citet{BMR91}. This is consistent with the results obtained by MOST.

As already noted, our dominant pulsation frequencies are consistent with those determined from the analysis
of the MOST data, but we find systematically different amplitude ratios. 
The MOST results indicate that $f_1$ shows the highest amplitude and $f_2$ is the
second highest: $A_1/A_2=1.1\pm0.1$ \citep{huber}.  However, in spectroscopy $f_2$
($P=11.931$~min) exhibits the highest amplitude for nearly all lines ($A_1/A_2=0.7$--0.9). One can speculate that this 
discrepancy comes from a different sampling of the propagation regions in spectroscopy and photometry,  respectively.

After improving the estimate for the three high-amplitude pulsation frequencies in 10\,Aql, we determined amplitudes and phases 
(expressed as 
fractional pulsation period)
for 65 selected lines
based on this solution and using $\cos{\{2 \pi [(t-t_0)/P - \varphi]\}}$
to fit the RV curves instead of the usual $\cos{\{2 \pi [(t-t_0)/P + \varphi]\}}$, which correlates
a larger phase to a later RV maximum. 
The reference time adopted for pulsation phase calculations is HJD=24453919.60523. The results
of the frequency analysis of individual lines in the data set containing all 5 UVES nights are reported in
Table~\ref{lines}. For all measured lines we also performed a period search
with the periodogram method, which allowed us to estimate the probability of the pulsation signal detection ($1-$\,False
Alarm Probability (FAP), see \citealt{horne}). With a few exceptions, the FAP is below $10^{-4}$.

Tables~\ref{lines} also gives the equivalent width of the pulsating lines. Notably, all REE lines with amplitudes significantly
exceeding 100~\ms\ are very weak (equivalent width $\sim$\,10~m\AA\ or less). Their analysis becomes possible thanks to a
high S/N ratio achieved in the time-resolved UVES spectra of 10\,Aql.

\section{Phase relations between photometry and spectroscopy}
\label{phase}

Our spectroscopic time-series were carried out simultaneously with the MOST
photometry. This gives us a unique possibility to derive directly the phase lag between the
photometric and spectroscopic pulsational variations, which provide useful constraints for subsequent
modelling of oscillations. Earlier attempts of similar simultaneous analysis of the
spectroscopic and photometric observations were carried out only for one roAp star  -- HD~24712 --
by \citet{MWWY88} using ground-based photometry and by \citet{ryab2007a} using MOST photometry.

The high S/N and spectral resolution of the present observations of 10\,Aql allow us to derive phase lags for individual spectral lines,
sampling different atmospheric layers. In order to minimize the influence of the higher (with respect to
the spectroscopic observations) point-to-point scatter of the
photometric observations, we have constructed an artificial light curve
based on the frequencies, amplitudes and phases of the 3 dominant
pulsation modes ($f_1$, $f_2$ and $f_3$) as derived by a multi-sine fit to the MOST data. In a next step the
artificial time-series were cross-correlated with the RV
observations of the individual spectral lines.
The time interval for the cross-correlation was chosen from plus to minus 11.93 minutes (the period of the
main photometric frequency, $f_1$), with an increment of 1 second.
We note that the significant modulation of the pulsation amplitude due to beating of several
frequencies helps to distinguish between a phase lag of, e.g., -0.2 and +0.8, 
which is impossible for a monoperiodic star.

The results of the cross-correlation analysis for the representative set of spectral lines are given
in Table\,\ref{most}, where we define the phase lag as the difference  between the RV maximum of a
given line and the photometric maximum, expressed in seconds or as a 
fractional main pulsation period
$P=11.93$ min. The error of the phase lag determination was estimated as follows. We added a
normally distributed random signal to both the spectroscopic and photometric data 
assuming that the noise amplitude corresponds to the observational error. For spectroscopy the error 
was derived from the RV determinations, and for photometry the MOST team recommended to use 2\% of the
signal. A phase lag was derived from these noisy data and repeated 200 times. The resulting standard
deviation was adopted as an error of the phase lag determination.

In contrast to HD\,24712, where the RV maxima for all lines preceded the photometric maximum, in
10\,Aql the maximum of brightness occurs prior to the RV maxima but the phase lag itself depends
strongly on the line considered. It is minimal ($\approx$\,40~s) for the Eu\ii\ 6645~\AA\ line
and gradually increases for other lines, reaching a maximum of 550~s in Tb\iii. The growth of the
phase lag from Eu to Tb lines has the same character as in HD\,24712. It probably traces the
outward propagation of the pulsation wave through the atmospheric layers. 

\begin{table}
\caption{Phase lags (in seconds and 
as a fractional main period
$P=11.93$ min) between the maxima of luminosity and RV variations of different chemical species
in the atmosphere of 10\,Aql. 
}
\label{most}
\begin{center}
\begin{tabular}{lccc}
\hline
Line    &  $\lambda$ (\AA) & \multicolumn{2}{c}{Phase lag}    \\
        &                  & seconds & fractional period\\
\hline
Eu\ii     & 6645 & $40\pm~8$  & $0.05\pm0.01$ \\
H\i       & 6563 & $58\pm12$  & $0.06\pm0.02$ \\
Sm\ii     & 5053 & $84\pm24$  & $0.12\pm0.03$ \\
Gd\ii     & 5092 & $127\pm45$ & $0.22\pm0.06$ \\
Dy\iii     & 5471 & $222\pm9$ & $0.29\pm0.01$ \\
Dy\iii     & 5556 & $236\pm18$& $0.31\pm0.03$ \\
Ce\ii     & 5274 & $212\pm14$ & $0.32\pm0.02$ \\
Dy\iii    & 5730 & $217\pm17$ & $0.32\pm0.02$ \\
Nd\iii    & 5851 & $260\pm25$ & $0.40\pm0.03$ \\
Nd\iii    & 5802 & $273\pm23$ & $0.40\pm0.03$ \\
Nd\iii    & 5845 & $312\pm17$ & $0.44\pm0.02$ \\
Nd\iii    & 5294 & $442\pm30$ & $0.59\pm0.04$ \\
Nd\iii    & 5102 & $444\pm43$ & $0.62\pm0.06$ \\
Pr\iii    & 5300 & $449\pm18$ & $0.70\pm0.03$ \\
Tb\iii    & 5505 & $559\pm~7$ & $0.77\pm0.01$ \\
\hline
\end{tabular}
\end{center}
\end{table}

\section{Pulsations in spectral lines}
\label{puls}
\subsection{Phase -- amplitude diagrams}

The presence of chemical stratification in the atmospheres of Ap stars offers a
unique possibility to resolve the vertical structure of pulsation modes.
However, the line-forming regions for REE are difficult to derive.
Atmospheric modelling  for such purposes should account for chemical
stratification, deviations from Local Thermodynamical Equilibrium (LTE),
magnetic field, and, eventually, pulsation effects on the shape and intensity
of spectral lines with substantial RV variability. The formidable difficulties of
such an elaborated theoretical approach explain why up to now Nd and Pr line
formation calculations accounting for NLTE and magnetic field effects in
chemically stratified atmosphere have been carried out only for two roAp stars,
$\gamma$~Equ and HD\,24712 (Mashonkina, Ryabchikova \& Ryabtsev
\citeyear{MRR05}; \citealt{RMR07}).

The assumption that the line intensity or equivalent width represents  a proxy
of the relative formation heights, frequently used in early roAp pulsation
studies (e.g. \citealt{KH98}), is only justified for the same element, respectively ion. 
It is now well established that stratification effects
are dominant in the atmospheres of cool Ap stars and, therefore, the formation
region of the weak lines of one element is not necessarily located deeper in the 
atmosphere than for strong lines of another element. Consequently, discussing
pulsation amplitude or phase versus line depth in context of pulsation modes can 
be grossly misleading.

\citet{ryab2007b} suggested a different approach to the pulsation tomography
problem based on the assumption of a continuous amplitude versus phase relation 
for an outward propagating magnetoacoustic wave. 
They proposed to use phase-amplitude diagrams
to diagnose the structure of pulsation modes based on the following arguments:
If one neglects surface chemical
inhomogeneity (which is a reasonable assumption for slow-rotating roAp stars),
then both phase and amplitude may provide information on the
relative line formation depth. For instance, a phase-independent growth of the
amplitudes measured for groups of lines belonging to different chemical species
corresponds to a standing wave 
in the layers where these chemical species are formed. In the case of a phase-dependent
amplitude trend, the lines showing later RV maximum should originate higher in
the atmosphere.
In the general case the vertical pulsation structure of roAp stars is likely to be a superposition
of standing and running waves, corresponding to the decoupled magnetic and acoustic pulsation
components which have different depth dependence \citep{SC07}. In the same star pulsations in certain ions may trace
the standing, magnetic part of the wave while other ions may show predominantly the running, acoustic component.
The phase-amplitude diagram method is well-suited to characterize this behaviour thus offering
the possibility of tracing the vertical properties of pulsation waves and studying the physics of magneto-acoustic oscillations
from trends of pulsation velocity amplitude and pulsation phase without tedious 
assignment of the physical depth to each pulsation measurement.

The key assumption used in interpreting the phase-amplitude diagrams is
that a later RV maximum corresponds to a higher line formation. This is supported by 
the gradual increase of pulsation phase with height from H$\beta$-, H$\alpha$-cores 
to Nd\ii-\iii\ -- Pr\ii-\iii\ lines \citep[e.g.,][]{SRB06} in practically all roAp stars where
formation heights of pulsating lines could be constrained independently of the pulsation analysis \citep{MRR05,RMR07}.
Since the chemical properties of most roAp stars are quite similar \citep{RNW04} and their 
phase-amplitude dependences are also comparable \citep{ryab2007b}, the assumption of continuous outward increase of pulsation phase
appears to be reasonable from the empirical point of view.
On the other hand, the general validity of this picture has been questioned in the theoretical
work by \citet{SC08}. These authors demonstrated that using a certain combination of the model parameters (not applicable to 10\,Aql or to
any other roAp star), one can obtain a complex superposition of standing and outwardly propagating waves which would mimic an inwardly 
running wave. This situation may complicate interpretation of the phase-amplitude diagrams,
and the case of 33\,Lib (Sect.\,\ref{33Lib}) possibly represents an example of such complex situation.
However, a more realistic theoretical modelling using pulsation and atmospheric parameters relevant for specific roAp stars
is required to confirm the reality of such complex depth dependence of pulsation phase.
 
\begin{figure}
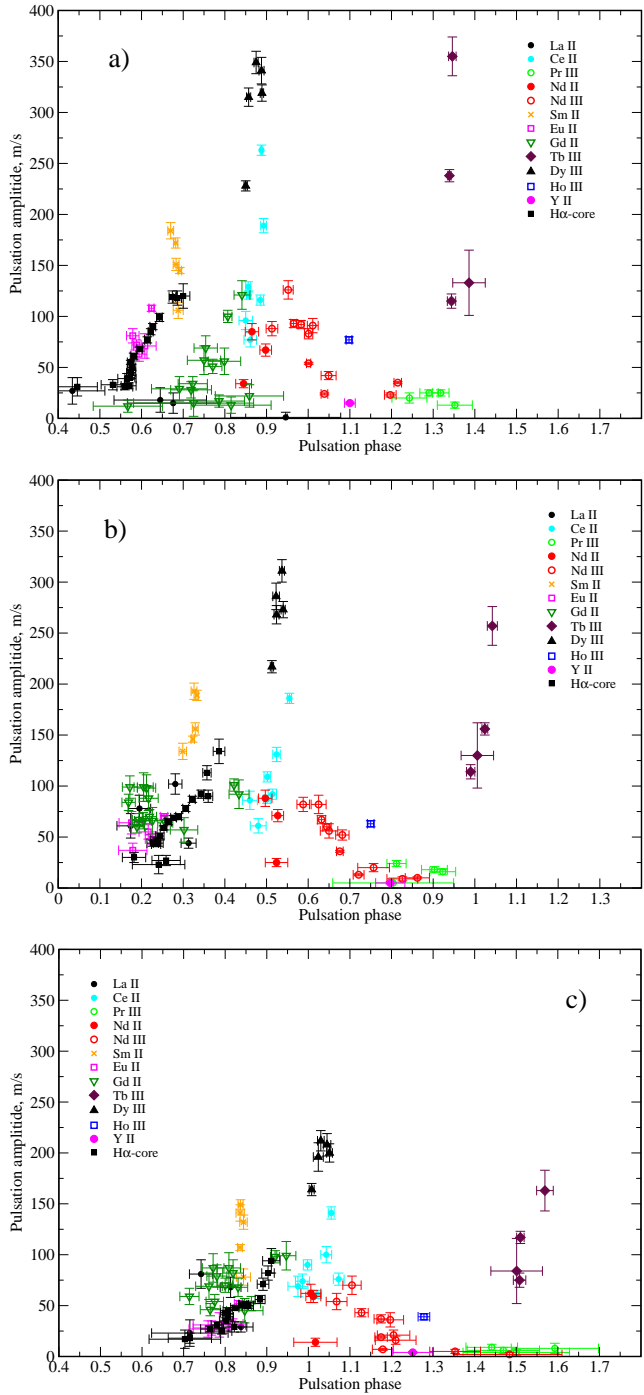

\centering
\figps{rv-ph_all_per1.eps} \\
\vspace*{0.2cm}
\figps{rv-ph_all_per2.eps} \\
\vspace*{0.2cm}
\figps{rv-ph_all_per3.eps}
\caption{Amplitude-phase diagrams for three main pulsation periods
in 10\,Aql: $P=11.93$ min (a), $P=11.51$ min (b), and $P=11.68$ min (c).}
\label{rv-ph}
\end{figure}

Similar to the analysis of 10 roAp stars presented by \citet{ryab2007b}, we
applied the amplitude-phase diagram method to 10\,Aql. Furthermore, 
the availability of our rich spectroscopic observational
material allows us to improve the precision of our previous analysis 
which was presented at the CP\#Ap Workshop in 2007 by \citet{SKR07},
and for the first time to recover the amplitude-phase diagrams for several pulsation
modes in the same roAp star. These diagrams are shown in Fig.~\ref{rv-ph}
for the 3 main frequencies observed in 10\,Aql.

For the purpose of comparison, we plot bisector measurements across the
H$\alpha$ core in the amplitude-phase diagrams constructed via centre-of-gravity
measurements. These measurements provide us with a physical depth
scale in the atmosphere because hydrogen is distributed homogeneously with
height. The core of the H$\alpha$ line is formed in the region between
$\log\tau_{5000}\sim-2$ and $\log\tau_{5000}\sim-4.7$ according to 
NLTE calculations using the codes described in \citet{MZG07},

As in other roAp stars, measurable pulsation amplitudes appear in the layers
where Eu\ii, La\ii, and the core of H$\alpha$ are formed, reach maximum and
then show decrease of the amplitude.  The phase 
increases by 0.2 
across the H$\alpha$-line core while the RV amplitude grows by 4--5 times. We
observe a running wave-like  behaviour in this part of the stellar atmosphere.
Above the formation height of the deepest point of the  H$\alpha$ profile
($r=0.15$) the amplitude continues to grow rapidly without significant phase
change. However, while in most other roAp stars the maximum RV amplitudes are
observed in singly and doubly ionised Nd and Pr lines, in 10\,Aql the maximum
are seen for the lines of Ce\ii, Dy\iii\ and a couple of unidentified lines too,
attributed by us to Dy\iii.

The lines of Nd and Pr are formed above the Dy\iii\ lines and show a RV
amplitude rapidly falling to very small values. In even higher layers, the pulsation
amplitude starts to grow again and reaches a new maximum in lines of Tb\iii.
Simultaneously, the pulsation phase changes by $\sim$0.5 
between the
two maxima. This is a typical picture of a pulsation node in the stellar atmosphere:
the phase changes by $180^{\circ}$ while  on both sides of the phase jump we
observe comparable RV amplitudes. The presence of the node is supported by the
bisector measurements in Nd\iii\ lines (see below). 
If the optical depth scale
indicated by the core of the H$\alpha$ line is valid for other lines too, then we
expect that the pulsation node is located near or above the optical depth of
$\log\tau_{5000}\sim-4.5$.
However, it should be stressed that the amplitude and
phase distributions with height in the atmospheres of roAp stars is expected to depend on the
local magnetic field strength and inclination \citep{SC07}. Thus, the inferred location of the node
observed in the disk-integrated quantities represents the average of the local vertical
pulsation structure which, possibly, exhibits substantial variation over the stellar surface.
Ideally one should take this variation into account when determining the exact vertical 
location of the node.

\begin{figure}
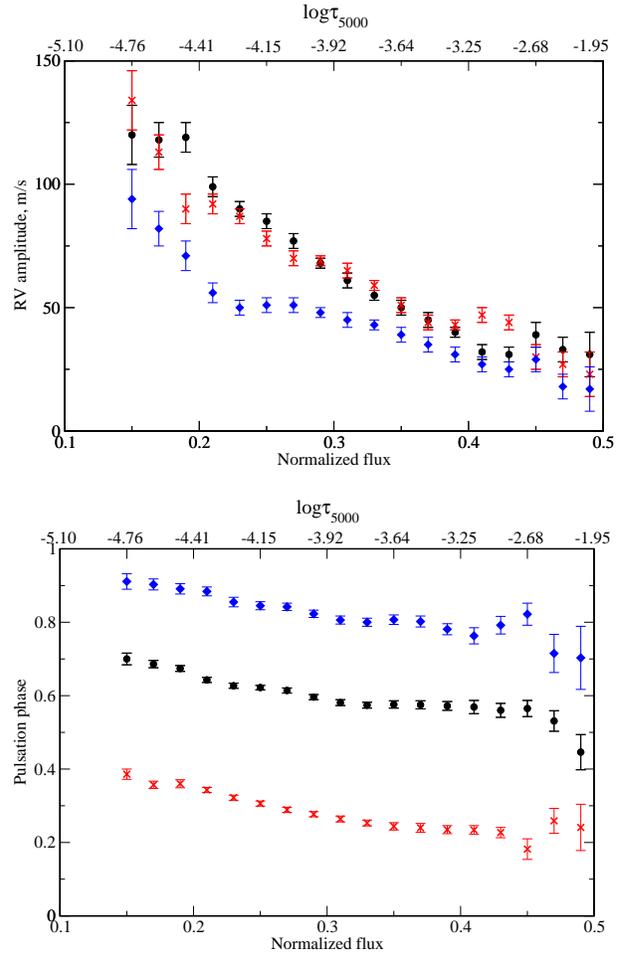

\centering
\fifps{80mm}{h_alpha_RV.eps}\\
\vspace*{0.4cm}
\fifps{80mm}{h_alpha_ph.eps}
\caption{Bisector RV amplitudes (top) and phases (bottom) across the
H$\alpha$ core for the three main pulsation periods in 10\,Aql:
$P_1 = 11.51$ min (crosses), $P_2 = 11.93$ min (filled circles), and $P_3 = 11.68$ min (diamonds).
The upper x-axis indicates the optical depth scale.}
\label{halpha}
\end{figure}

Among the lines of elements other than rare earths only the strongest line of
Y\ii, $\lambda$~5662.9~\AA, shows low-amplitude but definite pulsation. As in
most other roAp stars \citep{ryab2007b}, this line overlaps with Pr\iii\ lines
in the amplitude-phase diagrams.

The amplitude-phase diagrams for the three pulsation modes are very similar
except, perhaps, Gd\ii\ and La\ii\ lines. Only 4 out of 14 Gd\ii\ lines have
equivalent widths larger than 7 m\AA. Also, most of Gd\ii\ lines are shallow
due to substantial magnetic splitting; their central depth does not exceed 4--5
\%. This makes pulsation measurements difficult. Pulsations can be detected in
all Gd\ii, and it seems that this element shows higher amplitudes for the 11.51
and 11.68 min periods. The same situation is found for La\ii\ lines, which have
even lower equivalent widths and lower RV amplitudes. This discrepant variation
of the Gd\ii\ and La\ii\ lines for different frequencies may indicate  a
difference in the vertical extension of the corresponding modes. In particular, on
the basis of these measurements, one may suspect that the frequency $f_2$
($P=11.93$ min) corresponds to a pulsation mode located in higher atmospheric
layers compared to $f_1$ and $f_3$. This agrees with the difference between
photometric and spectroscopic amplitude ratios for  $f_1$ and $f_2$.

\subsection{Bisector variation}

The RV amplitudes and
phases across the H$\alpha$ core are shown in Fig.\,\ref{halpha} as a
function of the normalized flux (scale at the bottom) and optical depth (scale at the top) 
for all three principal pulsation
periods seen in 10\,Aql. We find very similar behaviour for the three modes.
In fact, the amplitudes of $f_1$ and $f_2$ are identical within error bars, whereas
the amplitude of $f_3$ is slightly lower. This suggests that the three
modes sample the hydrogen line core formation region in a similar manner.

Because of the weakness of most spectral lines, the only metal line suitable for a 
precise bisector analysis is Nd\iii~5102.43~\AA. This line is deep enough and
its wings are free from blends. In many roAp stars this Nd\iii\  line is
blended with Nd\ii~5102.39~\AA\ and should be used with caution. However, due
to the overall weakness of the Nd\ii\ spectrum in 10\,Aql, the blending effect is
negligible.

\begin{figure}
\centering
\includegraphics[width=85mm]{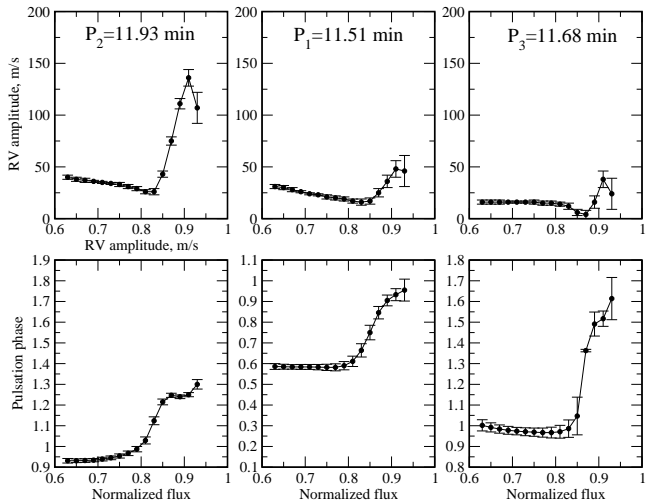}
\caption{
The RV amplitudes (top) and phases (bottom, expressed in as a fractional pulsation period) for
Nd\iii\ 5102~\AA.} \label{bis_Nd}
\end{figure}

Bisector measurements across the Nd\iii~5102~\AA\ line profile are shown in Fig.
\ref{bis_Nd}. A phase jump of approximately 0.4--0.5 
seems to be
located in similar atmospheric layers for all three main frequencies.
A bisector analysis provides an independent evidence for the pulsation node in
the upper atmosphere of 10\,Aql. This is the second case after 33\,Lib
(\citealt{MHK03}; Kurtz, Elkin \& Mathys \citeyear{KEM05}; \citealt{ryab2007b})
of a pulsation node found in a roAp star located within the REE-rich cloud.

Nd\iii~5102~\AA\ line provides the most clear signatures of the pulsation node: phase jump of $\sim$0.5 of the period
accompanied by the minimum in RV distribution. 
The start of the phase jump is seen in other Nd\iii\ lines as illustrated by Fig.\,6 of \citet{EKM08}.
In Nd\iii~6145\,\AA\ line it also reaches $\sim180^{\circ}$, however, we did not analyze this line due to its known
blend with Si\i\ line that influences bisector RV values. Another strong Nd\iii~6237\,\AA\ line was omitted from
the bisector analysis, too, because it is blended  in the red wing by one of Sm\ii, an element which also shows pulsation, 
but probably is formed deeper in the atmosphere (see our Fig.\,\ref{rv-ph}).  

In comparison to the bisector variation in the H$\alpha$ core, the Nd\iii~5102~\AA\ bisectors
behave differently for the three modes. The amplitude in the outer wings is the highest
for $f_2$, whereas the steepest phase jump is seen for $f_3$. 
Based on the location of the maximum change of the bisector phase further
outside
in the wings, we tentatively conclude that the node region is located somewhat 
deeper for the $f_3$ mode than for the other two frequencies.

\section{Nd\iii\ line profile asymmetry}
\label{profiles}

\begin{figure*}
\centering
\vskip 2mm
\includegraphics[width=13.5cm]{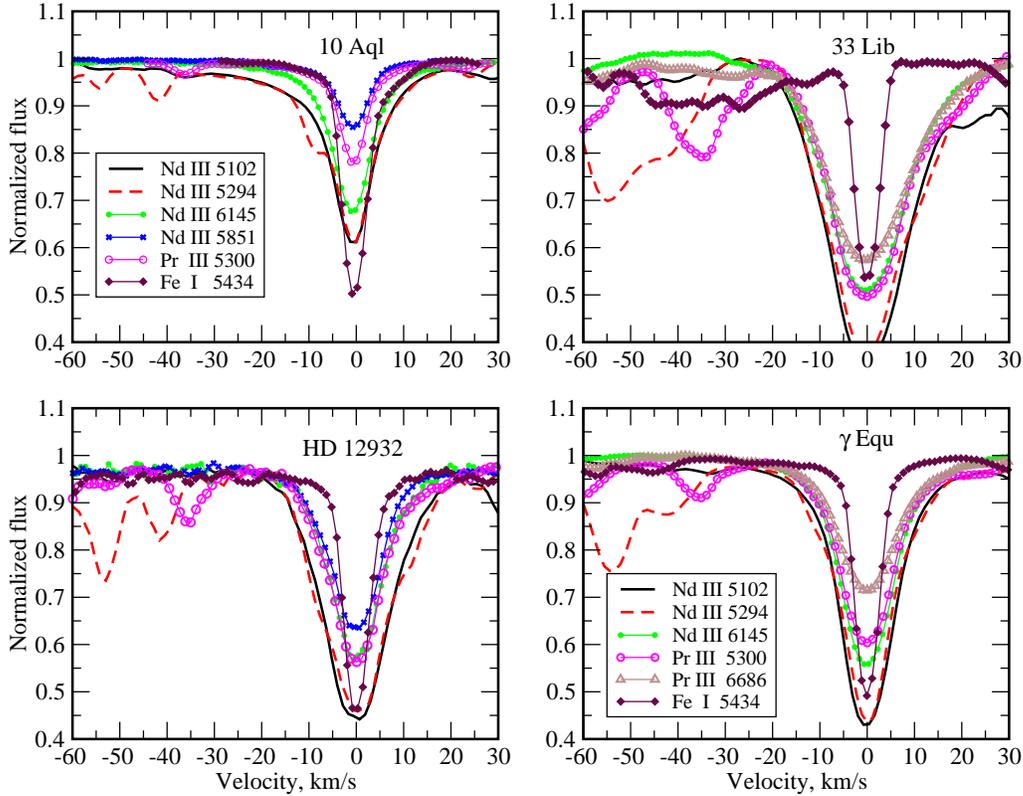}
\vskip 2mm
\caption{REE and Fe\i\ line profiles in the spectra of the weakly magnetic roAp stars 10\,Aql, 
and HD\,12923, and of stronger magnetic roAp stars 33\,Lib, and $\gamma$~Equ.} \label{LP}
\end{figure*}

Many cool Ap stars exhibit an unusually large broadening of the doubly ionized Nd and Pr lines, for
which macroturbulent velocities of the order of 10~\kms\ are often needed to achieve a good fit to
the observed line profiles \citep{KR01,ryab2007b}. The nature of this phenomenon is poorly
understood, but it could be related to a turbulent instability in the upper
atmospheric layers, possibly caused by the anomalous temperature gradient produced by a stratified
distribution of the chemical elements. Although roAp stars are the ones most frequently showing this
anomaly, it does not correlate with pulsation amplitude and is equally strong in stars in which
pulsations at the level of 20--30~\ms\ are barely detectable \citep{KRB08}. 
As discussed by \citet{KRW07} the modulation of turbulence by the periodic
contraction and expansion of the pulsating atmosphere in roAp stars with higher amplitudes 
explains the characteristic asymmetric line
profile variation of the REE lines, first detected by \citet{KR01} in $\gamma$~Equ.

10\,Aql, too, shows remarkably broad lines of Nd\iii, but curiously we find that the
strongest of these lines are also extremely asymmetric. This effect is illustrated in Fig.~\ref{LP}
for several Nd\iii\ lines. In this figure we show rare-earth line profiles in the spectra of four
slowly rotating roAp stars: two stars with a small magnetic field, 10\,Aql (\teff\,=\,7550~K,
\lgg\,=\,4.0, \vs\,=\,2.0~\kms, \bs\,=\,1.5~kG) and HD~12932 (\teff\,=\,7620~K, \lgg\,=\,4.2,
\vs\,=\,2.0~\kms, \bs\,=\,1.7~kG), and two stars with a large magnetic field, 33\,Lib
(\teff\,=\,7550~K, \lgg\,=\,4.3, \vs\,$\le$\,2.0~\kms, \bs\,=\,5~kG) and $\gamma$~Equ
(\teff\,=\,7700~K, \lgg\,=\,4.2, \vs\,$\le$\,1.0~\kms, \bs\,=\,4.1~kG) \citep{ryab2007b}. For a
comparison the profile of the magnetically insensitive non-pulsating (or weakly-pulsating)
Fe\i~5434 line is also shown for all stars. The wavelength scale is expressed in velocity units relative
to the line centres.  The strongest Nd\iii\ lines in 10\,Aql at $\lambda$ 5102 and 5294~\AA\ have a 
remarkable blue asymmetric wing reaching out to $-50$~\kms. The red wing stops at
$\approx$\,20~\kms\ as do both wings in the Nd\iii\ lines in the other roAp stars shown in Fig.\,\ref{LP}.
In 10\,Aql the blue wing asymmetry is still prominent in Nd\iii\ 6145~\AA\ and gradually disappears
for weaker Nd\iii\ lines. The profiles of strongly and weakly pulsating lines in HD\,12932
and $\gamma$\,Equ are symmetric. In 33\,Lib the weakly pulsating Fe\i~5434~\AA\ line is symmetric while
REE lines have a small red asymmetry. In 10\,Aql the non-pulsating Fe\i~5434~\AA\ line has a clearly
asymmetric red wing.

The available information is insufficient to attribute REE line asymmetries either to 
non-radial pulsations or to an anisotropic turbulent velocity field. The sign of the
asymmetry suggests an outward flow of Nd-rich material in the layers probed by the wings of strong
Nd\iii\ lines. However, this effect can be mimicked by a nonisotropic velocity distribution of an
ion stably stratified in the upper stellar atmosphere, as discussed by \citet{M78} for Sr in hotter
CP stars. A similar investigation is needed for REE in cool Ap stars to shed light on the 
abnormal shapes of strong Nd lines in 10\,Aql.

\section{Rotational modulation}

\begin{figure}
\figps{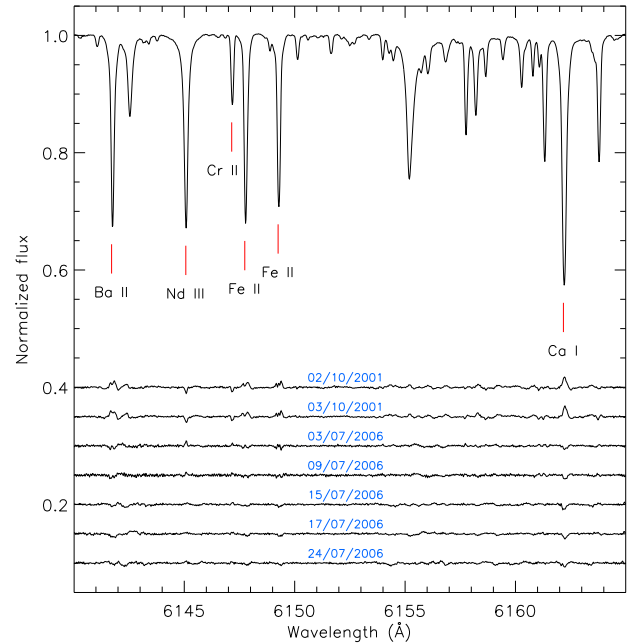}
\caption{Comparison between the 10\,Aql spectra obtained in 2001 and 2006. The upper plot shows
the average spectrum, the residual spectra are plotted below. They are shifted vertically
for a better visibility.} \label{longvar}
\end{figure}

A comparison of the average spectra obtained in five UVES nights over a period of
about 1 month did not show any hint of line profile or radial velocity changes on time scales of days. This supports
the study by \citet{RWA05} who have used own mean longitudinal field
measurements and data taken from the literature to argue for a long rotational period of 10\,Aql.
To search for long-term variations we have compared the average UVES spectra
with observations obtained in 2001 with the Gecko coud\'e spectrograph at the 3.6\,m
Canada-France-Hawaii telescope \citep{KLR02}. We introduced additional broadening to the latter
spectra to compensate for the difference in spectral resolution between Gecko and UVES.
The global average spectrum of 10\,Aql in the 6140--6165~\AA\ region and the residual spectra to this average 
are presented in Fig.\,\ref{longvar}.

A signature of weak variability can be seen in the centres of few strong lines of Ca\i, Fe\ii,
Ba\ii, Nd\iii, as well as in the moderately weak Cr\ii\ line. We are confident that this variability
is not an artifact due to using different spectrographs,  because the intensity variations of Ca\i,
Fe\ii, Ba\ii\ lines are in {\it antiphase} with the variations of Nd\iii\ and Cr\ii\ lines. Possibly, this is
another evidence for a rotational modulation in the spectra of 10\,Aql, indicating
a long rotation period.

\section{Discussion}
\label{discus}
\subsection{Reassessment of asteroseismic models}

\begin{figure}
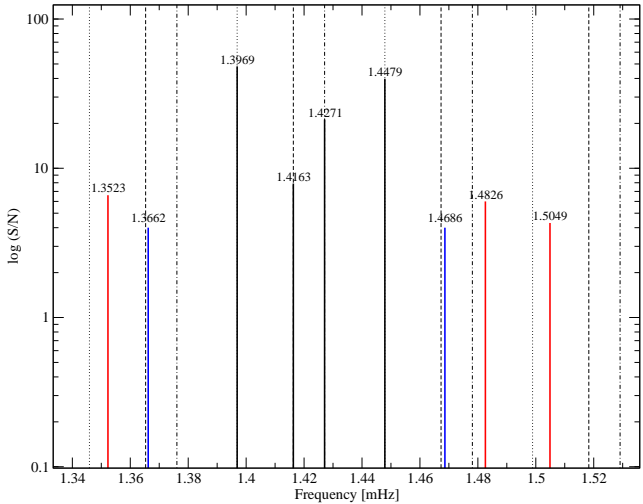

\figps{10_Aql_fig4.eps}
\caption{The observed secure (black lines) and tentative frequencies from the photometry (blue lines) and spectroscopy (red lines).
The expected position of radial orders for three different spherical degrees $\ell$ (vertical dashed, dotted and dashed-dotted lines, respectively)
is calculated based on a large separation of $f_{1}-f_{2}=51\,\mu$Hz.}
\label{spacing}
\end{figure}

The additional spectroscopic frequencies could offer a new view on the
large frequency separation $\Delta \nu$, a crucial factor for asteroseismology which is directly
connected to the mean density in the star and describes the separation of consecutive radial
overtones for high-order acoustic pulsation. Figure~\ref{spacing} shows a schematic amplitude
spectrum including all intrinsic and candidate frequencies of both MOST photometry and spectroscopy.
To first view, there is no apparent equal spacing visible. If we consider the four frequencies with
high S/N (solid lines) to estimate the spacing, the only reasonable solution corresponds to
$f_{1}-f_{2}=51\,\mu$Hz, a value already noted by \citet{huber}. Lower values (such as 20 or
30\,$\mu$Hz) although seem tempting, would however contradict the previously determined temperature
and luminosity of 10\,Aql (see Fig. 4 in \citealt{huber}). Assuming $\Delta\nu=51\,\mu$Hz to be
correct, Fig.\,\ref{spacing} shows the expected position of other radial orders for three different
spherical degrees $\ell$ (vertical dashed, dotted and dashed-dotted lines, respectively) based on
the position of the {\bf four highest} S/N peaks. Evidently, the agreement is not very convincing. Nevertheless,
an average deviation of about 5\,$\mu$Hz (which is needed to align the observed values to the
expected position) is about the order of the frequency shifts of consecutive radial overtones due to
the magnetic field perturbation as predicted by theory (see, e.g., \citealt{cunha2006}). It must be
noted, however, that such suggestion can lead to (almost) any desired result and therefore have to
be considered with extreme caution.

Generally, the frequencies we derived from spectroscopy do not contradict to those from MOST photometry. But
spectroscopic data provide an additional information useful for modelling. According to \citet{huber}
the best model describing the pulsation features in 10\,Aql derived from spectroscopy corresponds
to a star of 1.95$M_{\odot}$, \teff\,=\,7730 K and normal solar composition 
with the boundary reflection layer at $\log\tau_{5000}\sim-4$. The spectroscopic analysis reveals a
pulsation node in the Nd\ii\ - Nd\iii\ line formation region. A Non-LTE analysis
of these lines suggested that they are formed  in roAp stars in atmospheric layers above
$\log\tau_{5000}=-4$ \citep{MRR05}, thus indicating the position of the node. Indeed, a
pulsation node is predicted at $\log\tau_{5000}=-4$ in a star of 1.95$M_{\odot}$, \teff\,=\,7730 K but
assuming helium depletion (Saio, private communication), which would better correspond to the typical
chemistry of a magnetic peculiar star. Therefore, combining simultaneous photometric and spectroscopic 
observations, we expect to improve the pulsation model of
10\,Aql. This analysis will be presented in the next paper.

\subsection{Interpretation of spectroscopic pulsation measurements}

Here, as in many other recent studies of pulsational RV variations of roAp stars, we interpret the measurements of amplitude and
phase in terms of outward propagation of pulsation waves in a chemically stratified stellar atmosphere. Previously, this
interpretation of spectroscopic observations of multiperiodic roAp stars was handicapped by short spectroscopic
time series which did not allow to resolve frequency patterns of multiperiodic pulsators. 
Here we overcome this difficulty by combining time-resolved
observations acquired during several nights, but with a poor duty cycle, with a continuous photometric monitoring by MOST at 
the same time.  This strategy allows us to resolve the vertical structure of the three principal modes in 10\,Aql 
and to study the propagation of pulsation waves independently for all three frequencies.

We detect a radial node in the region sampled by the formation heights of Nd and Pr for all three
frequencies. The maximum amplitudes are observed for the \dy\ and Tb\iii\ lines that are presumably formed below and above
the region of the node. The existence of a node is supported by the bisector analysis of the \nd\ 5102~\AA\
line.

By attributing the differences in the RV curves of different REE ions to the vertical structure of oscillations we have
implicitly assumed that different species sample the horizontal structure of pulsation modes in a similar manner. This might
not be the case if the vertical distribution of pulsation amplitude and phase depends strongly on the stellar surface
position  due to a spotty element distribution combined with an intrinsic dependency of the mode structures on the orientation
and strength of the magnetic field. The importance of the latter effect, arising from a superposition of magnetic and
acoustic components of pulsation waves, has been recently emphasized by \citet{SC07}.

Due to a long rotation period we are unable to constrain the surface abundance distribution of 10\,Aql
in the same way as it could be done for rapidly rotating roAp stars \citep{K06x}. For this reason it is not straightforward to
distinguish between the vertical and horizontal structural effects. At the same time we see only small changes in the line profiles 
in a comparison of observations obtained in 2001 and 5 years later. 
Furthermore, the independent evidence for a node provided by phase-amplitude diagrams and bisector analyses
of different REE ions suggests that the observed phase variation corresponds rather to a vertical structure and not to a different
horizontal sampling of the stellar surface.

\subsection{Comparison with 33\,Lib}
\label{33Lib}

\begin{figure}
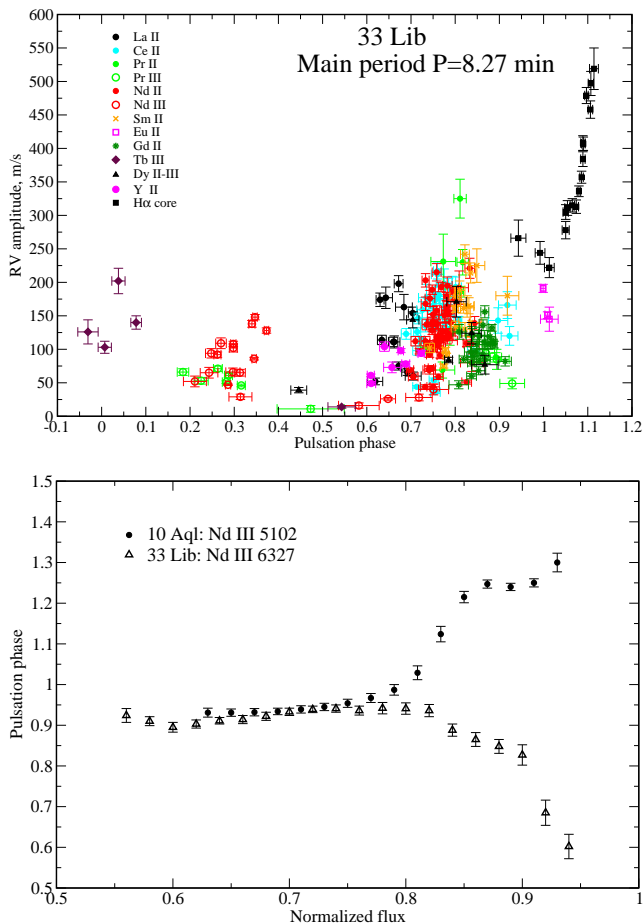

\figps{33Lib_ph-rv_all.eps}\\
\vspace*{-0.1cm}

\figps{10Aql-33Lib_bis1.eps}
\caption{Amplitude-phase diagram for 33\,Lib (upper panel) and pulsation phases of bisector measurements
at various normalized flux levels of \nd\ lines in 10\,Aql and 33\,Lib (lower panel). The phases are given 
as fractional pulsation period.
For demonstration purpose the phases of 33\,Lib (lower panel) are shifted by +0.6.}
\label{33Lib_ph-rv_all}
\end{figure}

A comparison of the pulsation properties of 33\,Lib and 10\,Aql is of considerable interest because
these are the only roAp stars with clear signatures of a radial node in the upper atmosphere.
\citet{ryab2007b} performed a pulsation analysis of 33\,Lib using phase-amplitude diagrams, which
enable a direct comparison of the vertical pulsation structure in two stars. The top panel of
Fig.\,\ref{33Lib_ph-rv_all} displays the phase-amplitude diagram for 33\,Lib, while
bisector measurements of \nd\ lines are compared for both stars in the bottom panel of
Fig.\,\ref{33Lib_ph-rv_all}.

A pulsation node is clearly present in the atmospheres of both stars but the phase jump has
opposite sign. 
Taking into account that the REE elements are concentrated in the upper atmospheric
layers close to or even above the H$\alpha$ line core formation zone, we can conclude from Figs.\,\ref{rv-ph} and
\ref{33Lib_ph-rv_all}  that the phase changes by 0.5 
between the formation zones
of Dy\iii\ and Tb\iii\ in the upper atmosphere of 10\,Aql, while it jumps by $-0.5$ between Nd\ii\
and Nd\iii\ in 33\,Lib.

Interestingly, the bisector measurements give us just the opposite phase variation with depth if one
treats the line profile as being produced in a normal stellar atmosphere where the line core is
formed higher than the line wings. As we show in the lower panel of Fig.~\ref{33Lib_ph-rv_all},
the phase decreases from the wing to the core in 10\,Aql, and increases in 33\,Lib. At present we do not
have a definite interpretation of this discrepant behaviour. One can suspect that the
the broad wings of strong REE lines actually originate in the outer atmosphere and hence show
pulsational characteristics of these layers.

Although the atmospheric parameters of 10\,Aql and 33\,Lib are similar, 33\,Lib has a much stronger
magnetic field, \bs\,=\,5.0~kG, and it has a higher overabundance of REEs (see \citealt{RNW04}).
In addition, it has a shorter main pulsation period, $P\,=\,8.27$~min, with the first harmonic
exhibiting the highest amplitude close to the position of a radial node. 
The negative phase jump and the shape of the amplitude-phase diagram (Fig.~\ref{33Lib_ph-rv_all}, upper panel) 
may be interpreted as a superposition of standing and running pulsation waves mimicing an inwardly propagating 
wave, as discussed by \citet{SC07}. 
A detailed study of chemical stratification and
atmospheric structure of both stars is required for a secure interpretation of pulsation results
and subsequent theoretical modelling.

\subsection{Comparison with previous spectroscopic pulsation studies of 10\,Aql}
\label{comp}

Non-radial pulsations in 10\,Aql were investigated in four different spectroscopic observing campaigns,  including the present
one. \citet{KLR02} found that RV amplitudes in Eu\ii\ and Gd\ii\ lines exceed those determined for Nd\iii\ line, which was not typical for
a roAp star. Additionally, \citet{KLR02} detected a change of RV amplitude between two consecutive nights of their observations. 
These characteristics of pulsational behaviour of 10\,Aql are confirmed in the present more extensive analysis.

\citet{HM05} observed the star during 3 nights, clearly detecting pulsational variability in 5 lines. They found the
highest pulsation amplitudes of a few hundred \ms\ in the three spectral lines of $\lambda\lambda$~5373, 5471, 5730\,\AA,  which could not
be identified. We also find significant amplitudes in these spectral features and tentatively identify these  lines as Dy\iii,
which allowed us to investigate variation of these lines in context of pulsation wave propagation through a stellar atmosphere
with chemical stratification.  

A recent study by \citet{EKM08} presented measurements of the pulsation amplitudes and phases based on UVES
spectroscopic time-series over two hours, which we included in our present analysis allowing for an increase of our spectroscopic data by 10\%.
\citet{EKM08} assumed that 10\,Aql pulsates with a main frequency of 1.428~mHz, which is, in fact, close to one of the
three main frequencies unambiguously detected in our spectroscopy and in the MOST photometry,  but with the lowest amplitude. 
The second frequency detected by \citet{EKM08} at 1.309~mHz is
based on variations of Tb\iii\ lines, and is not seen in our more voluminous data. It probably is an
artifact caused by, e.g., the limited time span of their spectroscopic observations of this multiperiodic pulsator.

Our study is the first to resolve the frequency spectrum of 10\,Aql with RV measurements. We find, in agreement with
precise space photometry, that pulsations of 10\,Aql are dominated by \textit{three} modes of comparable amplitude.  Based
on simulteneous spectroscopic and photometric observations, we were able to infer pulsational characteristics of individual 
modes as a function of height in the atmosphere of 10\,Aql. We also obtained phase shifts between luminosity and RV
variations providing constraints for modelling of roAp pulsation.     

Since our time-series analysis is based on a set of three frequencies, the RV amplitudes that we report for individual lines cannot
be directly compared with the results of previous studies which assumed monoperiodic variation of the star. However, a good
agreement with the results by \citet{EKM08} is found if we compare relative amplitudes and phase differences of the spectral
features in common. The two studies obtained qualitatively similar phase-amplitude diagrams, with the exception of discordant
behaviour of weak Er\iii~5903~\AA\ line studied by \citet{EKM08}. The pulsation characteristics of this line deviate from the
generally smooth phase-amplitude behaviour of other REE ions. We omitted this line from our pulsation analysis because it is
located in the spectral region full of telluric lines and is blended by 3 atmospheric lines of different intensity. This
blending is the most likely the reason for the deviating behaviour of the spectral line Er\iii~5903~\AA.

\subsection{Amplitude modulation in spectroscopy}
\label{concl}

Time-resolved spectroscopic observations of roAp stars seemed to indicate that the pulsation amplitudes of REE lines can be modulated on
a time scale of a few hours. This was first noted by \citet{KR01b} for $\alpha$\,Cir and later observed by \citet{KEM06} for
a few other roAp stars. Since in many cases this modulation cannot be linked to known photometric pulsation frequencies,
\citet{KEM06} speculated that the amplitude modulation in spectroscopy suggests the discovery of a new type of pulsational
behaviour in the upper atmospheres of roAp stars. However, using relatively short and typically 2-hour long time series resulting in only 100--150
individual measurements, this discovery is doubtful, because such
short observational data sets do not allow to resolve the frequency spectrum of multiperiodic roAp stars. In this context it
comes not as a surprise that for the only two roAp stars, HD\,24712 \citep{ryab2007a} and 10\,Aql (this paper), for
which an extensive spectroscopic (and photometric) monitoring over many nights was performed, 
no unexplained spectroscopic amplitude modulation could be found.
All amplitude modulations in 10\,Aql can be explained by beating effects of close frequencies. Hence, we recommend extreme caution in
the interpretation of amplitude changes seen in short data sets.

\section*{Acknowledgments}

We thank Dr. A. Ryabtsev for providing by us with the unpublished data on transition probability
calculations of Dy\iii, Dr. L. Mashonkina for her help in NLTE calculations of hydrogen lines and
Mag. T. Kallinger for his comments on phase lag determinations. In particular, we thank 
Dr. H. Saio for extremely useful discussion on pulsational modelling of 10~Aql atmosphere.  
Resources provided by the electronic databases (VALD, SIMBAD,
NASA's ADS) are acknowledged. This work was supported by the
Presidium RAS program, by research grants from the RFBI (08-02-00469a),
from the Swedish \textit{Kungliga Fysiografiska S\"allskapet} and
\textit{Royal Academy of Sciences} (grant No. 11630102), and  from the
Austrian Science Fund (FWF-P17580).

\label{lastpage}

\begin{thebibliography}{99}
\bibitem[\protect\citeauthoryear{Babcock}{1958}]{B58} Babcock H. W., 1958, ApJS, 3, 141
\bibitem[\protect\citeauthoryear{Belmonte et al.}{1991}]{BMR91}
Belmonte J. A., Martinez Roger C., \& Roca Cortes T., 1991, A\&A, 248, 541
\bibitem[\protect\citeauthoryear{Belmonte et al.}{1992}]{BKM92}
Belmonte J. A., Kreidl T. J., \& Martinez Roger C., 1992, IBVS, 3752
\bibitem[\protect\citeauthoryear{Bigot \& Dziembowski}{2002}]{BD02}
Bigot L., \& Dziembowski W. A., 2002, A\&A, 391, 235
\bibitem[\protect\citeauthoryear{Cunha}{2006}]{cunha2006}
Cunha M. S., 2006, Mem. Soc. Astr. It., 77, 447
\bibitem[\protect\citeauthoryear{Cunha et al.}{2003}]{CFM03}
Cunha M. S., Fernandes J. M. M. B., \& Monteiro M. J. P. F. G., 2003, A\&A, 343, 831
\bibitem[\protect\citeauthoryear{Elkin et al.}{2008}]{EKM08}
Elkin V. G., Kurtz D. W., \& Mathys, G., 2008, MNRAS, 386, 481
\bibitem[\protect\citeauthoryear{Hatzes \& Mkrtichian}{2005}]{HM05}
Hatzes A. P., \& Mkrtichian D. E., 2005, A\&A, 430, 279
\bibitem[\protect\citeauthoryear{Heller \& Kramer}{1988}]{HK88}
Heller C. H., \& Kramer K. S., 1988, PASP, 100, 583
\bibitem[\protect\citeauthoryear{Heller \& Kramer}{1990}]{HK90}
Heller C. H., \& Kramer K. S., 1990, MNRAS, 244, 372
\bibitem[\protect\citeauthoryear{Horne \& Baliunas}{1986}]{horne}
Horne J.H., \& Baliunas S.L., 1986, ApJ, 302, 757
\bibitem[\protect\citeauthoryear{Huber et al.}{2008}]{huber}
Huber D., Saio H., Gruberbauer M., et al., 2008, A\&A, 483, 239
\bibitem[\protect\citeauthoryear{Kanaan \& Hatzes}{1998}]{KH98}
Kanaan A., \& Hatzes A. P., 1998, ApJ, 503, 848
\bibitem[\protect\citeauthoryear{Kochukhov \& Ryabchikova}{2001a}]{KR01}
Kochukhov O., \& Ryabchikova T., 2001, A\&A, 374, 615
\bibitem[\protect\citeauthoryear{Kochukhov \& Ryabchikova}{2001b}]{KR01b}
Kochukhov O., \& Ryabchikova T., 2001, A\&A, 377, L22
\bibitem[\protect\citeauthoryear{Kochukhov et al.}{2002}]{KLR02}
Kochukhov O., Landstreet J. D., Ryabchikova T. A., Weiss W. W., \& Kupka F., 2002, MNRAS, 337, L1
\bibitem[\protect\citeauthoryear{Kochukhov}{2004a}]{K04a}
Kochukhov, O., 2004a, in \textit{The A-star puzzle}, eds. J.\,Zverko, J.\,\v{Z}i\v{z}\v{n}ovsk\'{y},
S.J.\,Adelman, W.W.\,Weiss, Cambridge University Press, IAUS~224, 433
\bibitem[\protect\citeauthoryear{Kochukhov}{2004b}]{K04b}
Kochukhov O., 2004b, ApJ, 615, L149
\bibitem[\protect\citeauthoryear{Kochukhov}{2005}]{K05}
Kochukhov O., 2005, in
{\it Element Stratification in Stars: 40 Years of Atomic Diffusion},
eds. G. Alecian, O. Richard and S. Vauclair, EAS Publ. Ser., 17, 103
\bibitem[\protect\citeauthoryear{Kochukhov}{2006}]{K06x}
Kochukhov O., 2006, A\&A, 446, 1051
\bibitem[\protect\citeauthoryear{Kochukhov \& Bagnulo}{2006}]{KB06}
Kochukhov O., \& Bagnulo S. 2006, A\&A, 450, 763
\bibitem[\protect\citeauthoryear{Kochukhov}{2007a}]{K07a}
Kochukhov O., 2007a, in
{\it Vienna Workshop on the Future of Asteroseismology}, eds. G. Handler and G. Houdek,
Comm. in Asteroseismology, 150, 39
\bibitem[\protect\citeauthoryear{Kochukhov}{2007b}]{K06}
Kochukhov O., 2007b, in \textit{Physics of Magnetic Stars},
eds. I.I. Romanyuk and D.O. Kudryavtsev, Nizhnij Arkhyz, 109
\bibitem[\protect\citeauthoryear{Kochukhov et al.}{2007}]{KRW07}
Kochukhov O., Ryabchikova T., Weiss W.W., Landstreet J.D., \& Lyashko D., 2008, MNRAS, 376, 651
\bibitem[\protect\citeauthoryear{Kochukhov et al.}{2008}]{KRB08}
Kochukhov O., Ryabchikova T., Bagnulo S., \& Lo Curto G., 2008, in
{\it CP\#Ap Workshop}, eds. J.\,\v{Z}i\v{z}\v{n}ovsk\'{y}, J. Zverko, E. Paunzen, M. Netopil,
Contrib. Astr. Obs. Skalnat\'e Pleso, 38, 423 (astro-ph/0711.4923)
\bibitem[\protect\citeauthoryear{Kurtz}{1982}]{k1982}
Kurtz D. W., 1982, IBVS, 1436, 1
\bibitem[\protect\citeauthoryear{Kurtz \& Martinez}{2000}]{KM00}
Kurtz D. W., \& Martinez P., 2000, Baltic Astron., 9, 253
\bibitem[\protect\citeauthoryear{Kurtz et al.}{2005}]{KEM05}
Kurtz D. W., Elkin V. G., \& Mathys G., 2005, MNRAS, 358, L6
\bibitem[\protect\citeauthoryear{Kurtz et al.}{2006}]{KEM06}
Kurtz D. W., Elkin V. G., \& Mathys G., 2006, MNRAS, 370, 1274
\bibitem[\protect\citeauthoryear{Kupka et al.}{1999}]{vald2}
Kupka F., Piskunov N., Ryabchikova T. A., Stempels H. C., \& Weiss, W. W., 1999, A\&As, 138, 119
\bibitem[\protect\citeauthoryear{Lenz \& Breger}{2004}]{period04}
Lenz P., \& Breger M., 2004, in \textit{The A-star puzzle}, eds. J.\,Zverko, J.\,\v{Z}i\v{z}\v{n}ovsk\'{y},
S.J.\,Adelman, W.W.\,Weiss, Cambridge University Press, IAUS~224, 786
\bibitem[\protect\citeauthoryear{Leone et al.}{2003}]{LVS03}
Leone F., Vacca W.D., \& Stift M.J., 2003, A\&A, 409, 1055
\bibitem[\protect\citeauthoryear{Leone \& Catanzaro}{2004}]{LC04}
Leone F. \& Catanzaro G., 2004, A\&A, 425, 271
\bibitem[\protect\citeauthoryear{Lyashko et al.}{2007}]{lyashko}
Lyashko D. A., Tsymbal V. V., \& Makaganuik V. A., 2007, in
\textit{Spectroscopic methods in modern astrophysics}, eds. L. Mashonkina and M. Sachkov, Moscow, 100
\bibitem[\protect\citeauthoryear{Martinez et al.}{1994}]{MSH94}
Martinez P., Sekiguchi K., \& Hashimoto O., 1994, MNRAS, 268, 169
\bibitem[\protect\citeauthoryear{Mashonkina et al.}{2005}]{MRR05}
Mashonkina L., Ryabchikova T., \& Ryabtsev A., 2005, A\&A, 441, 309
\bibitem[\protect\citeauthoryear{Mashonkina et al.}{2007}]{MZG07}
Mashonkina L., Zhao G,. Geren T., et al., 2008, A\&A, 478, 529
\bibitem[\protect\citeauthoryear{Matthews et al.}{1988}]{MWWY88}
Matthews J. M., Wehlau W. H., Walker G. A. H., \& Yang S., 1988, ApJ, 324, 1099
\bibitem[\protect\citeauthoryear{Matthews et al.}{1996}]{MWR96}
Matthews J. M., Wehlau W. H., Rice J., \& Walker G. A. H., 1996, ApJ, 459, 278
\bibitem[\protect\citeauthoryear{Matthews et al.}{1999}]{MKM99}
Matthews J. M., Kurtz D. W., \& Martinez P., 1999, ApJ, 511, 422
\bibitem[\protect\citeauthoryear{Michaud}{1978}]{M78}
Michaud G., 1978, ApJ, 220, 592
\bibitem[\protect\citeauthoryear{Mkrtichian et al.}{2003}]{MHK03}
Mkrtichian D. E., Hatzes A. P., \& Kanaan A., 2003, MNRAS, 345, 781
\bibitem[\protect\citeauthoryear{Nesvacil et al.}{2008}]{N08}
Nesvacil N., Weiss W. W., \& Kochukhov O., 2008,
{\it CP\#Ap Workshop}, eds. J.\,\v{Z}i\v{z}\v{n}ovsk\'{y}, J. Zverko, E. Paunzen, M. Netopil,
Contrib. Astr. Obs. Skalnat\'e Pleso, 38, 329
\bibitem[\protect\citeauthoryear{Piskunov}{1999}]{P99}
Piskunov N. E., 1999, in \textit{2nd International Workshop on
Solar Polarization}, eds. K. Nagendra and J. Stenflo, Kluwer Acad. Publ. ASSL, 243, 515
\bibitem[\protect\citeauthoryear{Preston}{1970}]{P70}
Preston G. W., 1970, PASP, 82, 87
\bibitem[\protect\citeauthoryear{Reegan}{2007}]{sigspec}
Reegan P., 2007, arXiv:physics/0703160
\bibitem[\protect\citeauthoryear{Ryabchikova}{2004}]{R04}
Ryabchikova T., 2004, in \textit{The A-star puzzle}, eds. J.\,Zverko, J.\,\v{Z}i\v{z}\v{n}ovsk\'{y},
S.J.\,Adelman, W.W.\,Weiss, Cambridge University Press, IAUS~224, 283
\bibitem[\protect\citeauthoryear{Ryabchikova et al.}{2000}]{RSH00}
Ryabchikova T. A., Savanov I. S., Hatzes A. P., Weiss W. W., \& Handler G., 2000, A\&A, 357, 981
\bibitem[\protect\citeauthoryear{Ryabchikova et al.}{2004}]{RNW04}
Ryabchikova T., Nesvacil N., Weiss W. W, Kochukhov O., \& St\"utz Ch., 2004, A\&A, 423, 705
\bibitem[\protect\citeauthoryear{Ryabchikova et al.}{2005}]{RWA05}
Ryabchikova T., Wade G. A., Auri\'re M., et al., 2005, A\&A, 429, L55
\bibitem[\protect\citeauthoryear{Ryabchikova et al.}{2006}]{RRKB06}
Ryabchikova T., Ryabtsev A., Kochukhov O., \& Bagnulo S., 2006, A\&A, 456, 329
\bibitem[\protect\citeauthoryear{Ryabchikova et al.}{2007a}]{ryab2007a}
Ryabchikova T., Sachkov M., Weiss W.W., et al., 2007a, A\&A, 462, 1103
\bibitem[\protect\citeauthoryear{Ryabchikova et al.}{2007b}]{ryab2007b}
Ryabchikova T., Sachkov M., Kochukhov O., \& Lyashko D., 2007b, A\&A, 473, 907
\bibitem[\protect\citeauthoryear{Ryabchikova et al.}{2007c}]{RMR07}
Ryabchikova T., Mashonkina L., Ryabtsev A., Kildiyarova R., \& Khristoforova M., 2007c, CoAst., 150, 81
\bibitem[\protect\citeauthoryear{Ryabchikova et al.}{2008}]{RKB08}
Ryabchikova T., Kochukhov O., \& Bagnulo S., 2008, A\&A, 480, 811
\bibitem[\protect\citeauthoryear{Sachkov et al.}{2006}]{SRB06}
Sachkov M., Ryabchikova T., Bagnulo S., et al., 2006, Comm. in Asteroseismology, 147, 97
\bibitem[\protect\citeauthoryear{Sachkov et al.}{2007}]{SKR07}
Sachkov M., Kochukhov O., Ryabchikova T., Leone F., Bagnulo S., Weiss W.W., 2008,
in {\it CP\#Ap Workshop}, eds. J.\,\v{Z}i\v{z}\v{n}ovsk\'{y}, J. Zverko, E. Paunzen, M. Netopil,
Contrib. Astr. Obs. Skalnat\'e Pleso, 38, 323 (astro-ph/0712.1340)
\bibitem[\protect\citeauthoryear{Saio \& Gautschy}{2004}]{SG04}
Saio H., \& Gautschy A., 2004, MNRAS, 350, 485
\bibitem[\protect\citeauthoryear{Saio}{2005}]{S05}
Saio H. 2005, MNRAS, 350, 1022
\bibitem[\protect\citeauthoryear{Savanov et al.}{1999}]{SMR99}
Savanov I. S., Malanushenko V. P., \& Ryabchikova T. R., 1999, Astron. Lett., 25, 802
\bibitem[\protect\citeauthoryear{Sousa \& Cunha}{2008a}]{SC08}
Sousa S.G., \& Cunha M.S., 2008a, MNRAS, 386, 351
\bibitem[\protect\citeauthoryear{Sousa \& Cunha}{2008b}]{SC07}
Sousa J.C., \& Cunha M.S., 2008b, 
{\it CP\#Ap Workshop}, eds. J.\,\v{Z}i\v{z}\v{n}ovsk\'{y}, J. Zverko, E. Paunzen, M. Netopil,
Contrib. Astr. Obs. Skalnat\'e Pleso, 38, 453 (astro-ph/0712.2973)
\bibitem[\protect\citeauthoryear{Tsymbal et al.}{2003}]{tsymbal}
Tsymbal V., Lyashko D., \& Weiss W. W., 2003, in {\it Modelling of Stellar Atmospheres}, IAU Symp. No. 210, eds. N. Piskunov,
W.W. Weiss, D.F. Gray, ASP, E49
\bibitem[\protect\citeauthoryear{Walker et al.}{2003}]{W2003}
Walker G., Matthews G., Kuschnig R., et al., 2003, PASP, 115, 1023

\end{thebibliography}
\end{document}